\documentclass[authoryear,3p,11pt]{elsarticle}
\usepackage{amsmath,amssymb,mathtools,mathrsfs}

\usepackage{ulem}
\usepackage{cancel}
\usepackage{float}
\usepackage{soul}
\usepackage[round]{natbib}  
\usepackage{hyperref}
\hypersetup{hidelinks,
	colorlinks=true,
	allcolors=black,
	pdfstartview=Fit,
	breaklinks=true}

\usepackage[usenames,dvipsnames]{xcolor}
\usepackage{booktabs}
\usepackage{multirow}
\usepackage{amsthm}  

\newtheorem{definition}{Definition}

\usepackage[symbol]{footmisc}

\journal{Elsevier}

\begin{document}

\begin{frontmatter}

\title{Inverse Elastica: A Theoretical Framework for Inverse Design of Morphing Slender Structures}


 \author{JiaHao Li$^{1,\dagger}$, Weicheng Huang$^{2,\dagger}$, YinBo Zhu$^{1}$, Luxia Yu$^{1}$, Xiaohao Sun$^{1,}$\footnote[1]{\textit{Corresponding Author: sunxiaohao@ustc.edu.cn (X.S.)}}, Mingchao Liu$^{3,}$\footnote[1]{\textit{Corresponding Author: m.liu.2@bham.ac.uk (M.L.)}}, HengAn Wu$^{1,}$\footnote[1]{\textit{Corresponding Author: wuha@ustc.edu.cn (H.W.)}}}


\address{$^{1}$\:CAS Key Laboratory of Mechanical Behavior and Design of Materials, Department of Modern Mechanics, University of Science and Technology of China, Hefei 230027, People’s Republic of China\\
$^{2}$\:School of Engineering, Newcastle University, Newcastle upon Tyne NE1 7RU, UK \\
$^{3}$\:Department of Mechanical Engineering, University of Birmingham, Birmingham B15 2TT, UK \\
$^{\dagger}$These authors contributed equally to this work.
}


\begin{abstract}

Inverse design of morphing slender structures with programmable curvature has significant applications in various engineering fields. Most existing studies formulate it as an optimization problem, which requires repeatedly solving the forward equations to identify optimal designs. Such methods, however, are computationally intensive and often susceptible to local minima issues. In contrast, solving the inverse problem theoretically, which can bypass the need for optimizations, is highly efficient yet remains challenging, particularly for cases involving arbitrary boundary conditions (BCs). Here, we develop a systematic theoretical framework, termed inverse elastica, for the direct determination of the undeformed configuration from a target deformed shape along with prescribed BCs. Building upon the classical elastica, inverse elastica is derived by supplementing the geometric equations of undeformed configurations.  The framework shows three key features: reduced nonlinearity, solution multiplicity, and inverse loading. These principles are demonstrated through two representative models: an analytical solution for a two-dimensional arc and a numerical continuation study of the inverse loading of a three-dimensional helical spring. Furthermore, we develop a theory-assisted optimization strategy for cases in which the constrains of the undeformed configurations cannot be directly formulated as BCs. Using this strategy, we achieve rational inverse design of complex spatial curves and curve-discretized surfaces with varying Gaussian curvatures. Our theoretical predictions are validated through both discrete elastic rod simulations and experiments. While grounded in theory, the engineering value of inverse elastica is demonstrated through design of a deployable and conformable hemispherical helical antenna. This work thus provides a novel strategy for the inverse design of morphing slender structures, opening new avenues for applications in morphing structures, soft robotics, deployable radio-frequency systems, architectural design, and beyond.

\end{abstract}


\end{frontmatter}

\section{Introduction}
\label{sec:Introduction}

Morphing structures that adapt their shape in response to external stimuli have gained importance due to applications in flexible electronics~\citep{xu2015assembly,fan2020inverse,liu2020tapered}, aerospace~\citep{benvenuto2015dynamics,huang2023contact}, biomedical devices~\citep{kim2019ferromagnetic,tong2025inverse}, and soft robotics~\citep{shin2018hygrobot,tong2025inverse}. These systems typically use low-dimensional geometries embedded in three-dimensional (3D) space, such as two-dimensional (2D) plates/shells and one-dimensional (1D) beams/ribbons.
In the past two decades, significant research has focused on 2D morphing structures. For example, \cite{efrati2009elastic} developed the non-Euclidean plate theory to model bilayer plate deformations, and further applied to capture helicoid-to-helix transitions in chiral seed pods~\citep{armon2011geometry}. It further shows that this bilayer can be approximated by a monolayer model~\citep{van2017growth}. Building on these theories, inverse design has been demonstrated using inflatable plates~\citep{siefert2019bio}, 4D printing~\citep{sydney2016biomimetic}, and liquid crystal elastomers~\citep{aharoni2018universal,boley2019shape}.
However, the design of 2D morphing structures is limited by Gauss’s Theorema Egregium, which states that Gaussian curvature depends solely on the surface metric~\citep{siefert2019bio}. Thus, achieving large-curvature target shapes from flat plates requires significant in-plane stretching, often beyond what material expansion can provide, restricting achievable morphologies.

\begin{figure}[ht]
    \centering
    \includegraphics[width=1.0\columnwidth]{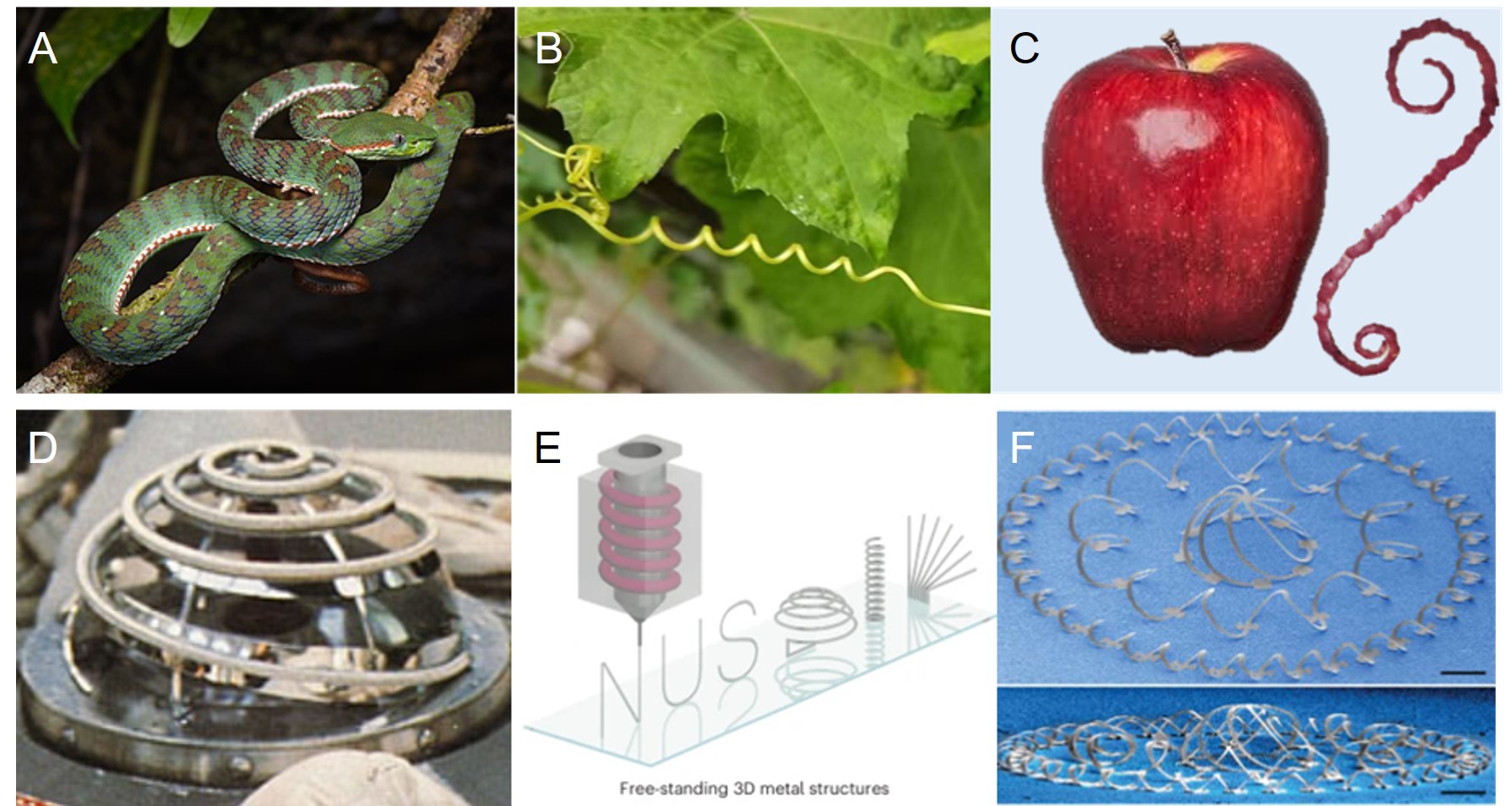}
    \caption{\textbf{Morphological diversity of slender structures in nature and corresponding engineering applications.}
    (A) Arboreal adaptation demonstrated by a Trimeresurus sabahi coiling around a tree branch~\citep{rushen2019viper}. (B)  A plant tendril with helix morphology under gravity.  (C) A continuous helical apple peel is generated during paring. (D)  \href{http://mentallandscape.com/V_Telemetry.htm}{The conformable hemispherical helix antenna in Venera-7.} (E) Freestanding three-dimensional metal slender structures fabricated via advanced additive manufacturing technology~\citep{ling2024tension}. (F) Flexible electronic self-assembly through buckling of slender structures~\citep{xu2015assembly}. }
    \label{fig:f1}
\end{figure}

To address this challenge, a promising approach is to reduce the structural dimensionality. Compared with 2D morphing structures, 1D morphing structures, often called morphing slender structures, offer greater geometric flexibility, which significantly expands the design space for inverse design. In nature, slender structures such as the Trimeresurus sabahi snake~\citep{rushen2019viper} and plant tendrils show evolutionary advantages due to their ability to adaptively change shape, enabling reliable attachment to branches and other surfaces, as illustrated in Figs.~\ref{fig:f1}(A) and (B).
Furthermore, ribbons represent an intermediate form between 1D and 2D structures. They combine the geometric flexibility of slender structures with the ability to discretize curved surfaces, effectively bypassing the constraints imposed by Gauss’s Theorema Egregium~\citep{li2025biomimetic}. This principle is exemplified by the continuous helical peel of an apple, shown in Fig.~\ref{fig:f1}(C).
The geometric flexibility of slender structures holds profound significance not only in the evolutionary adaptations of organisms but also in a wide range of engineering applications. Notably, the folded hemispherical helix plays a critical role in the design of compact antennas, with its quality factor approaching the fundamental Chu–Harrington limit~\citep{chu1948physical,kong2016electrically}. An exemplary application is the antenna used in the Venera-7 mission, as illustrated in Fig.~\ref{fig:f1}(D). Building on this, substantial advancements have been achieved in 3D printing technologies, enabling the fabrication of complex slender structures without the need for support materials, as demonstrated in Fig.~\ref{fig:f1}(E)~\citep{ling2024tension}. In addition to these advanced printing techniques, the morphology of slender structures formed through elastic deformation represents a relatively simple yet effective strategy, which is a widely applied strategy in flexible electronic fabrication as illustrated in Fig.~\ref{fig:f1}(F)~\citep{xu2015assembly,fan2020inverse,cheng2023programming}.

Morphing slender structures can be broadly classified into two categories based on their fundamental actuation mechanisms. The first category comprises structures actuated by external physical stimuli such as temperature gradients~\citep{timoshenko1925analysis}, hydration~\citep{armon2011geometry,chen2011tunable,shin2018hygrobot,li2025biomimetic}, or pH variations~\citep{sawa2011shape}. These materials often employ bilayer architectures, where strain mismatch between dissimilar layers induces bending and twisting deformations~\citep{armon2011geometry,chen2011tunable,li2025biomimetic,li2025harnessing}. Over the past century, the inverse design of such bilayer systems has been extensively explored, progressing from simple beam models to uniform deformations in helical ribbons~\citep{timoshenko1925analysis,armon2011geometry,chen2011tunable}, and more recently culminating in a comprehensive theoretical framework for complex spatial curves with various curvature and torsion~\citep{li2025biomimetic}. Despite the maturity of the design theory for morphing bilayer ribbons, experimental fabrication remains challenging, particularly in precisely encoding parameters such as the volume fraction of swelling component and the fibril orientation angle within the material.
The second category consists of morphing slender structures that are more readily fabricated experimentally by regulating elastic deformation. Designing the deformed configuration (DC) typically involves tuning the stiffness distribution or intrinsic curvatures of the undeformed configuration (UC) as primary design parameters. Manipulating stiffness distribution is a common approach in soft electronics; however, it is limited by a restricted design space and often requires complex boundary loading conditions to achieve effective morphing~\citep{fan2020inverse,liu2020tapered,zhang2022shape,kansara2023inverse}. For example, previous studies have discretized rotating bodies with varying Gaussian curvatures into multiple radially arranged beams with tailored width profiles. Morphing toward the target shape is realized by precisely controlling the Euler buckling of each beam, which necessitates simultaneous regulation of boundary displacements across all beams~\citep{fan2020inverse,liu2020tapered,yang2023morphing}. To overcome the challenges posed by such complex loading requirements, a more robust and comprehensive theoretical framework for inverse design of morphing slender structures is essential.

Regulating the intrinsic curvature of the UC enables a more straightforward inverse design process, as it typically requires only simple boundary conditions. Prior studies have drawn inspiration from natural deformation processes, such as the bending of curling hair or tendrils with intrinsic growth curvature under gravity, where one end of the rod is clamped and the other remains free. As a result, clamped-free boundary conditions are commonly adopted in inverse design theories. Bertails-Descoubes et al. investigated the inverse design of suspended Kirchhoff rods under gravity, both theoretically and numerically~\citep{derouet2010stable, derouet2013inverse, bertails2018inverse}. They demonstrated that a unique inverse solution exists under clamped-free boundary conditions. Subsequent works by \cite{qin2022bottom} and \cite{tong2025inverse} explored the two-dimensional cases.  
Although inverse design under gravity-driven clamped-free conditions has been studied, its applicability remains restricted to a narrow class of materials. This limitation arises because the gravity–elastic parameter must lie within a narrow range for the solution to be physically realizable~\citep{bertails2018inverse,qin2022bottom,tong2025inverse}. Moreover, most existing studies formulate inverse design as an optimization problem, in which the discrepancy between forward-computed and target configurations is minimized~\citep{derouet2010stable, derouet2013inverse, qin2020genetic}. However, such optimization-based approaches often incur excessive computational costs, because the underlying physical mechanism of inverse design remains unclear.  
This motivates a broader challenge: how to establish a general inverse theory that transcends the limitations of gravity and enables the inverse design of morphing slender structures under arbitrary boundary conditions?  
At its core, this question reflects a fundamental inversion of the classical elastica problem. The Kirchhoff rod equations, as formulated in classical elastica theory~\citep{love1944treatise,audoly2000elasticity,o2017modeling}, are used to determine the DC from a known UC. In contrast, inverse design based on intrinsic curvature aims to reconstruct the UC from a desired DC. Addressing this challenge necessitates the development of a general inverse elastica theory.

In this work, we develop an inverse theory for classical elastica theory, termed the inverse elastica, which serves as a dual counterpart to the classical Kirchhoff–Clebsch rod theory~\citep{love1944treatise,audoly2000elasticity,o2017modeling}. As a unified theoretical framework, the inverse elastica  enable the direct determination of the UC from the DC under arbitrary boundary conditions and external loads. We begin with a brief review of the forward problem in Section~\ref{sec:Forward problem}, followed by the formulation and theoretical development of the inverse problem in Section~\ref{sec: Inverse problem}. In analogy with the forward loading process that deforms the UC into the DC, we define an inverse loading that recovers the UC from the DC based on the proposed theory. To illustrate the three key features of the inverse elastica theory, namely reduced nonlinearity, multiplicity of solutions, and the concept of inverse loading, we present two illustrative toy models in Section~\ref{sec:model demo}. These include the inverse design of an arc and the inverse design of a helix, where both analytical solutions and numerical continuation methods are employed. Because certain constraints on the UC morphology, such as minimum curvature or minimum volume occupancy, are not easily expressed as explicit boundary conditions, we develop a theory-assisted optimization strategy for inverse design in Section~\ref{sec:opt}. Section~\ref{sec:examples} extends the framework to the inverse design of complex spatial curves, such as the trefoil knot, and helix discretized curved surfaces with varying Gaussian curvatures, including the torus, cone, sphere, and hyperboloid. Furthermore, in Section~\ref{sec:engineering} we demonstrate the potential engineering application of the framework through the inverse design of a deployable and conformable hemispherical antenna. Finally, Section~\ref{sec:dis} discusses the main features of the inverse elastica theory and outlines potential directions for future research. Conclusions are summarized in Section~\ref{sec:conclusion}.

\section{The theoretical framework for inverse design}
\label{sec: framework}
\begin{figure}[ht]
    \centering
    \includegraphics[width=1.0\columnwidth]{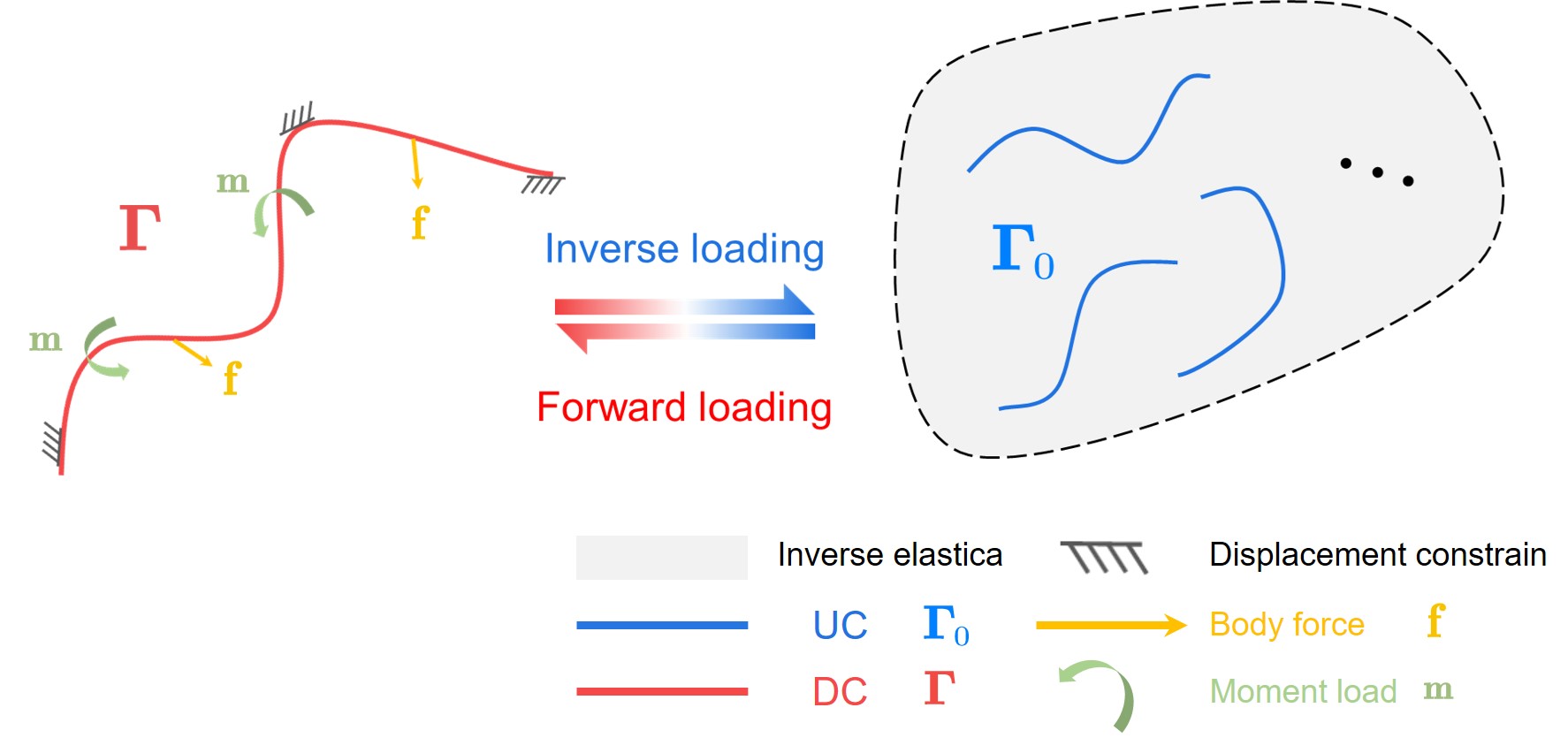}
    \caption{\textbf{Inverse elastica and elastica under arbitrary displacement boundary conditions and external loading.} The red curve indicates the DC, and the blue curve is the UC. A specific DC corresponds to a family of UC determined by the applied body forces, moment loads, and displacement constraints.}
    \label{fig:f2}
\end{figure}

In this section, we discuss the forward and inverse frameworks of morphing slender structures based on elastica theory. We first review the classical forward elastica theory for slender structures, then develop the inverse elastica theory.

\subsection{Forward elastica problem}
\label{sec:Forward problem}

\paragraph{Governing equations} The forward problem is defined as determining the DC of slender structures based on a known UC, material properties, applied boundary displacements, and loading conditions.
We use the classical Kirchhoff model to describe the mechanics of a slender rod and ribbon.
The Kirchhoff model, which accounts for bending and twisting while neglecting stretching, shearing, and cross-sectional deformation, has been widely used to predict the deformation behavior of slender structures under specified boundary conditions~\citep{audoly2000elasticity,o2017modeling}.
The Kirchhoff theory can serve as a solution to the forward problem.
As shown in Fig.~\ref{fig:f2}, the DC is denoted as \( \boldsymbol{\Gamma}(s) \), and the UC is denoted as \( \boldsymbol{\Gamma}_0(s) \), where $s \in [0,L]$ is the arc length parameter, and $L$ is the total length of the slender system.
We use a subscript with zero $ (\cdot)_{0 }$ for the evaluation on the UC.
Based on the force balance and the moment balance, the equilibrium equation is given by
\begin{equation}
\mathbf{F}'(s)+\mathbf{f}(s)=\boldsymbol{0},
\label{eqn:1}
\end{equation} 
\begin{equation}
\mathbf{M}'(s)+\mathbf{d}_3(s) \times \mathbf{F}(s)+\mathbf{m}(s)=\boldsymbol{0},
\label{eqn:2}
\end{equation} 
where $\mathbf{f}(s)$ and $\mathbf{m}(s)$ are the external force and moment density loads, $\mathbf{F}(s)$, $\mathbf{M}(s)$ are the internal force and internal moment of the slender system, $(\cdot)'=\partial(\cdot)/\partial s$ is the derivative of variable $( \cdot )$ to arc length parameter $s$, and $\mathbf{d}_3(s)$ is the tangent of the rod. Later, we ignore the arc length parameter $(s)$ for simplification.
To solve the above ordinary differential equations (ODEs), we expand all physical vectors in the local deformed material frame (DC), $\mathbf{D}=\{\mathbf{d}_1,\mathbf{d}_2,\mathbf{d}_3\}^T$, (where $\mathbf{d}_1$ and $\mathbf{d}_2$ are the two material directors), the equilibrium equation and the linear constitutive equation are given as
\begin{equation}
    \bar{\mathbf{F}}'+\bar{\mathbf{F}}\mathbf{\Omega }+\bar{\mathbf{f}}=0,
    \label{eqn:3}
\end{equation}
\begin{equation}
   \bar{\mathbf{M}}'+\bar{\mathbf{M}}\mathbf{\Omega }+\bar{\mathbf{F}}\mathbf{\Lambda }+\bar{\mathbf{m}}=0,
   \label{eqn:4}
\end{equation}
\begin{equation}
\bar{\mathbf{M}}=\left( \boldsymbol{\omega }-\boldsymbol{\omega }_0 \right) \mathbf{S},
\label{eqn:5}
\end{equation}
\begin{equation}
    \boldsymbol{\Lambda}=\left( \begin{matrix}
	0&		1&		0\\
	-1&		0&		0\\
	0&		0&		0\\
\end{matrix} \right), 
\label{eqn:6}
\end{equation}
where $\boldsymbol{\Omega} = \rm{skew} (\boldsymbol{\omega})$ is the rotation tensor, $\boldsymbol{\omega}=\{\omega_1,\omega_2,\omega_3\}$ is the Darboux vector in the DC, $\boldsymbol{\omega}_0$ is the Darboux vector in the UC, $\mathbf{S}=\rm{diag}(\{EI_1,EI_2,GJ\})$ is the stiffness matrix. Also, the physical vectors for the force and moment are transformed into the component vector, i.e., $\bar{\mathbf{F}}=\{F_1,F_2,F_3\}$ is the component vector of internal force vector $\mathbf{F}$ in material frame thus $\mathbf{F}=\bar{\mathbf{F}}{\mathbf{D}}$, and, similar for $\bar{\mathbf{M}}=\{M_1,M_2,M_3\}$ (and $\mathbf{M}=\bar{\mathbf{M}}\mathbf{D}$),
$\bar{\mathbf{f}}=\{f_1,f_2,f_3\}$ (and $\mathbf{f}=\bar{\mathbf{f}}\mathbf{D}$),
and $\bar{\mathbf{m}}=\{m_1,m_2,m_3\}$ (and $\mathbf{m}=\bar{\mathbf{m}}\mathbf{D}$).

\paragraph{Boundary conditions} To include the displacement boundary conditions, the rotation relationship in 3D space can be obtained as
\begin{equation}
\mathbf{\Gamma }'=\boldsymbol{R}_{3}^{T},
\label{eqn:7}
\end{equation}
\begin{equation}
    \boldsymbol{q}'=\mathbf{Q} \boldsymbol{q},
\label{eqn:8}
\end{equation}
where, $\boldsymbol{\Gamma}=\{x,\ y,\ z\}^{T}$ is the DC, $\boldsymbol{q}=\{q_0,q_1,q_2,q_3\}^T$ is the quaternion vectors, $\boldsymbol{R}_3$ is the third row of rotation matrix $\boldsymbol{R}$ and $\boldsymbol{\rm{Q}}$ is a matrix,
\begin{equation}
    \boldsymbol{R}=2\left( \begin{matrix}
	q_{0}^{2}+q_{1}^{2}-1/2&		q_1q_2+q_0q_3&		q_1q_3-q_0q_2\\
	q_1q_2-q_0q_3&		q_{0}^{2}+q_{2}^{2}-1/2&		q_2q_3+q_0q_1\\
	q_1q_3+q_0q_2&		q_2q_3-q_0q_1&		q_{0}^{2}+q_{3}^{2}-1/2\\
\end{matrix} \right),
\label{eqn:9}
\end{equation}
and
\begin{equation}
    \mathbf{Q}=\left( \begin{matrix}
	0&		-\boldsymbol{\omega }\\
	\boldsymbol{\omega }^T&		\mathbf{\Omega }\\
\end{matrix} \right).
\label{eqn:10}
\end{equation}
Combine Eqs.~\eqref{eqn:3} $\sim$~\eqref{eqn:8}, we can solve the forward problem as a boundary value problem (BVP) under the given displacement boundary condition of boundary displacement $\boldsymbol{\Gamma}(0)$, $\boldsymbol{\Gamma}(L)$ and rotation angle $\boldsymbol{q}(0)$, $\boldsymbol{q}(L)$, where $L$ is the length of the rod.

In this section, we define the forward problem of slender structures. Fundamentally, the forward problem involves determining the DC from a given UC, together with the applied loads and displacement boundary conditions. By using the elastica theory the forward problem can be solved completely with kirchhoff model. Note that here the process we concern is the inverse problem: determining the UC from DC, together with the applied loads and displacement constraints. In the next section, we will build an inverse elastica theory to solve the inverse problem completely.

\subsection{Inverse elastica theory}
\label{sec: Inverse problem}

The inverse problem is defined as determining the UC  with the given DC, material properties, applied loading and boundary displacement. We here establish the theoretical framework and develop the control equation for the inverse problem. Note that for the inverse process solving UC from DC, the solution is also required to meet the force and moment balance in DC, which means the solution should be self-consistent with the forward theory. Therefore, the physical equation of inverse elastica theory can be read from the Kirchhoff equation directly
\begin{equation}
    \begin{cases}
	\bar{\mathbf{F}}'+\bar{\mathbf{F}}\mathbf{\Omega }+\bar{\mathbf{f}}=\boldsymbol{0}\\
	\bar{\mathbf{M}}'+\bar{\mathbf{M}}\mathbf{\Omega }+\bar{\mathbf{F}}\mathbf{\Lambda }+\boldsymbol{m}=\boldsymbol{0}\\
	\bar{\mathbf{M}}=\left( \boldsymbol{\omega }-\boldsymbol{\omega }_0 \right) \mathbf{S}
\end{cases},
\label{eqn:11}
\end{equation}
Note that for the forward problem, $\bar{\mathbf{F}}$, $\boldsymbol{\omega}$, $\boldsymbol{\Gamma}$ and $\boldsymbol{q}$ are unknown and they can be solved by Eqs.~\eqref{eqn:3} $\sim$~\eqref{eqn:8} with the certain boundary condition. However, for the inverse problem the unknown is $\boldsymbol{\omega}_0$ rather than $\boldsymbol{\omega}$, and this means that to some extent, the nonlinearity of the equation has been reduced because $\boldsymbol{\Omega}$ as a known is no longer coupled with $\bar{\mathbf{M}}$ and $\bar{\mathbf{F}}$ in Eq.~\eqref{eqn:11}. For the clamped-free boundary condition, the solution to Eq.~\eqref{eqn:11} is unique. In fact, this result is intuitive because the clamped-free boundary condition enforces zero force and moment at the free end, i.e., $\bar{\mathbf{F}}(L) =\mathbf{0},\bar{\mathbf{M}}(L) = \mathbf{0}$ where $L$ denotes the total length of the rod. Consequently, Eq.~\eqref{eqn:11} effectively reduces to an initial value problem (IVP), and the boundary condition at the free end fully determine the system's response. By the well-known existence and uniqueness theorem for ODEs, the solution is guaranteed to be unique provided the governing equations satisfy the necessary smoothness and Lipschitz continuity conditions. For the more general case, if the force and moment at the end of the rod are known, for example, if an object with a known weight is attached at the end, and not just the clamped-free boundary condition, the solution to the inverse equation is also unique, because the initial values of forces and moments have been determined. 

Here our goal is to establish a general inverse elastica theory, without restricting to specific boundary conditions. As discussed in previous study, although it has been realized the solution to clamped-clamped boundary condition is not unique~\citep{bertails2018inverse}, a general theory to achieve the inverse design is still challenging. Suppose that we give a random initial value of $\bar{\mathbf{F}}(L)$ and $\bar{\mathbf{M}}(L)$, the solution $\boldsymbol{\omega}_0$ is a certain UC,  which can morph to DC by elastic deformation with the given force and moment applied to the right end. This means that, under displacement boundary condition (e.g., clamped–clamped), there exists a family of UC corresponding to a single DC, as illustrated in Fig.~\ref{fig:f2}. This is because a clamped boundary can support arbitrary magnitudes of force and moment. For the pinned boundary condition, the analysis is similar, except that the force $\boldsymbol{F}(L)$ can be arbitrary, while the moment $\bar{\mathbf{M}}(L)$ must be zero. For the case of multiple displacement boundary conditions as in a multi-point boundary value problem, the system can be simplified to a set of coupled BVPs, as shown in previous studies~\citep{yu2021numerical,sun2022phase,yu2023continuous, sun}. Therefore, in this work, we primarily consider the clamped–clamped boundary condition.
It is important to note that our objective is to control the displacement of the right end, rather than the force and moment. As a result, an additional geometric constraint equation for the UC is required. Since the DC is known, the relationship among geometric rotation, the centerline coordinates, and material curvature is inherently satisfied. Consequently, the geometric equations (Eqs.~\eqref{eqn:7} $\sim$~\eqref{eqn:8}) make no sense in the inverse elastica theory. However, it inspires us that the requirement that the centerline and the intrinsic material curvature of UC also must satisfy the geometric rotation constraint. Accordingly, We introduce the geometric constraint equation required for the inverse formulation 
\begin{equation}
    \mathbf{\Gamma }_0'=\boldsymbol{R}_{03}^{T},
    \label{eqn:12}
\end{equation}
\begin{equation}
    	\boldsymbol{q}_0'=\mathbf{Q} _0\boldsymbol{q}_0.
    \label{eqn:13}
\end{equation}
Similarly as Eqs.~\eqref{eqn:7} $\sim$~\eqref{eqn:8}, where $\boldsymbol{\Gamma}_0=\{x_0(s),y_0(s),z_0(s)\}^{T}$ is coordination of center line of UC as a function of arc length parameter $s$, $\boldsymbol{q}_0=\{q_{00}(s),q_{01}(s),q_{02}(s),q_{03}(s)\}^T$ is the quaternion vector of UC, $\boldsymbol{R}_{03}$ is the third row of rotation matrix $\boldsymbol{R}_0$ and $\boldsymbol{\rm{Q}}_0$ is a quaternion matrix for UC. $\boldsymbol{R}_0$ and $\boldsymbol{\rm{Q}}_0$ can be written as

\begin{equation}
   \boldsymbol{R}_0=2\left( \begin{matrix}
	q_{00}^{2}+q_{01}^{2}-1/2&		q_{01}q_{02}+q_{00}q_{03}&		q_{01}q_{03}-q_{00}q_{02}\\
	q_{01}q_{02}-q_{00}q_{03}&		q_{00}^{2}+q_{02}^{2}-1/2&		q_{02}q_{03}+q_{00}q_{01}\\
	q_{01}q_{03}+q_{00}q_{02}&		q_{02}q_{03}-q_{00}q_{01}&		q_{00}^{2}+q_{03}^{2}-1/2\\
\end{matrix} \right), 
    \label{eqn:14}
\end{equation}
and
\begin{equation}
    \mathbf{Q}_0=\left( \begin{matrix}
	0&		-\boldsymbol{\omega }_0\\
	\boldsymbol{\omega }_0^T&		\mathbf{\Omega }_0\\
\end{matrix} \right). 
    \label{eqn:15}
\end{equation}
Now we derived the complete inverse elastica theory for morphing slender structures
\begin{equation}
\begin{cases}
	\bar{\mathbf{F}}'+\bar{\mathbf{F}}\mathbf{\Omega }+\bar{\mathbf{f}}=\mathbf{0}\\
	\bar{\mathbf{M}}'+\bar{\mathbf{M}}\mathbf{\Omega }+\bar{\mathbf{F}}\mathbf{\Lambda }+\bar{\mathbf{m}}=\boldsymbol{0}\\
	\bar{\mathbf{M}}=\left( \boldsymbol{\omega }-\boldsymbol{\omega }_0 \right) \mathbf{S}\\
	\left( \mathbf{\Gamma }_0 \right) '=\boldsymbol{R}_{03}^{T}\\
	\left( \boldsymbol{q}_0 \right) '=\mathbf{Q}_0\boldsymbol{q}_0\\
\end{cases},
    \label{eqn:16}
\end{equation}

For the inverse problem, we can solve $\bar{\mathbf{F}}$, $\bar{\mathbf{M}}$, $\boldsymbol{{\Gamma}}_0$ and $\boldsymbol{q}_0$ from Eq.~\eqref{eqn:16} with the given displacement boundary condition $\boldsymbol{q}_0(0)$, $\boldsymbol{q}_0(L)$ and $\boldsymbol{\Gamma}_0(0)$, $\boldsymbol{\Gamma}_0(L)$ for UC. This is similar as the forward loading process in the forward problem, thus it was defined as inverse loading. We emphasize that forward loading and inverse loading are different, because the inverse theory can not be derived by simply swapping $\boldsymbol{\Gamma}_0, \boldsymbol{q}_0$ and $\boldsymbol{\Gamma}, \boldsymbol{q}$ in the Kirchhoff rod equation. The most important difference between them is that $\bar{\mathbf{F}}$ and $\bar{\mathbf{M}}$ are coupled with unknown rotation $\boldsymbol{\Omega}$ in forward elastica theory but the coupled $\boldsymbol{\Omega}$ is known in inverse elastica theory, which means the underlaying physics of inverse and forward loading is different. Intrinsically, the forward theory requires that forces and moments achieve equilibrium within the unknown configuration, whereas the inverse theory necessitates such equilibrium within the known configuration. Note that if we expand the balance equations of force and moment in the material frame of UC, we can also induce the unknown rotation $\boldsymbol{\Omega}_0$ of UC. However, under this situation the linear constitutive also needs to be expanded in the material frame of UC, which will bring the unnecessary nonlinearity in constitutive. 

Three key points are summarized as follows to clearly understand the inverse elastica theory :
\begin{itemize}
    \item \textbf{Reduced nonlinearity:} Compared with the forward theory, $\bar{\mathbf{F}}$, $\bar{\mathbf{M}}$ is no longer coupled with the unknown $\boldsymbol{\Omega}$ in inverse theory, which reduces the nonlinearity of the control equation.
    \item \textbf{Multiplicity of solutions:} UC is unique when the $\bar{\mathbf{F}}$, $\bar{\mathbf{M}}$ are known at one end, but not unique for displacement boundary condition (pin or clamped) because the arbitrary value of $\bar{\mathbf{F}}$, $\bar{\mathbf{M}}$ can be supplied. 
    \item \textbf{Inverse loading:} UC can be obtained from DC by inverse loading, which is a similar process as forward loading but transforming DC to UC.
\end{itemize}
To comprehend the three key characteristics of inverse elastica theory clearly, we introduce two illustrative models in the following section: the analytical inverse design of a planar arc and the inverse loading of a helix with numerical continuation.

\subsection{Two illustrative models}
\label{sec:model demo}

In this subsection, we find the UCs of two illustrative models by applying the inverse elastica theory. During this process, we not only find the analytical solution for 2D inverse elastica problem but also use the numerical continuation method to compare the inverse loading with forward loading. The following two examples provide a concrete understanding of the three key concepts discussed in the previous section.

\subsubsection{2D case: Inverse design of an arc}

\begin{figure}[ht]
    \centering
    \includegraphics[width=1.0\columnwidth]{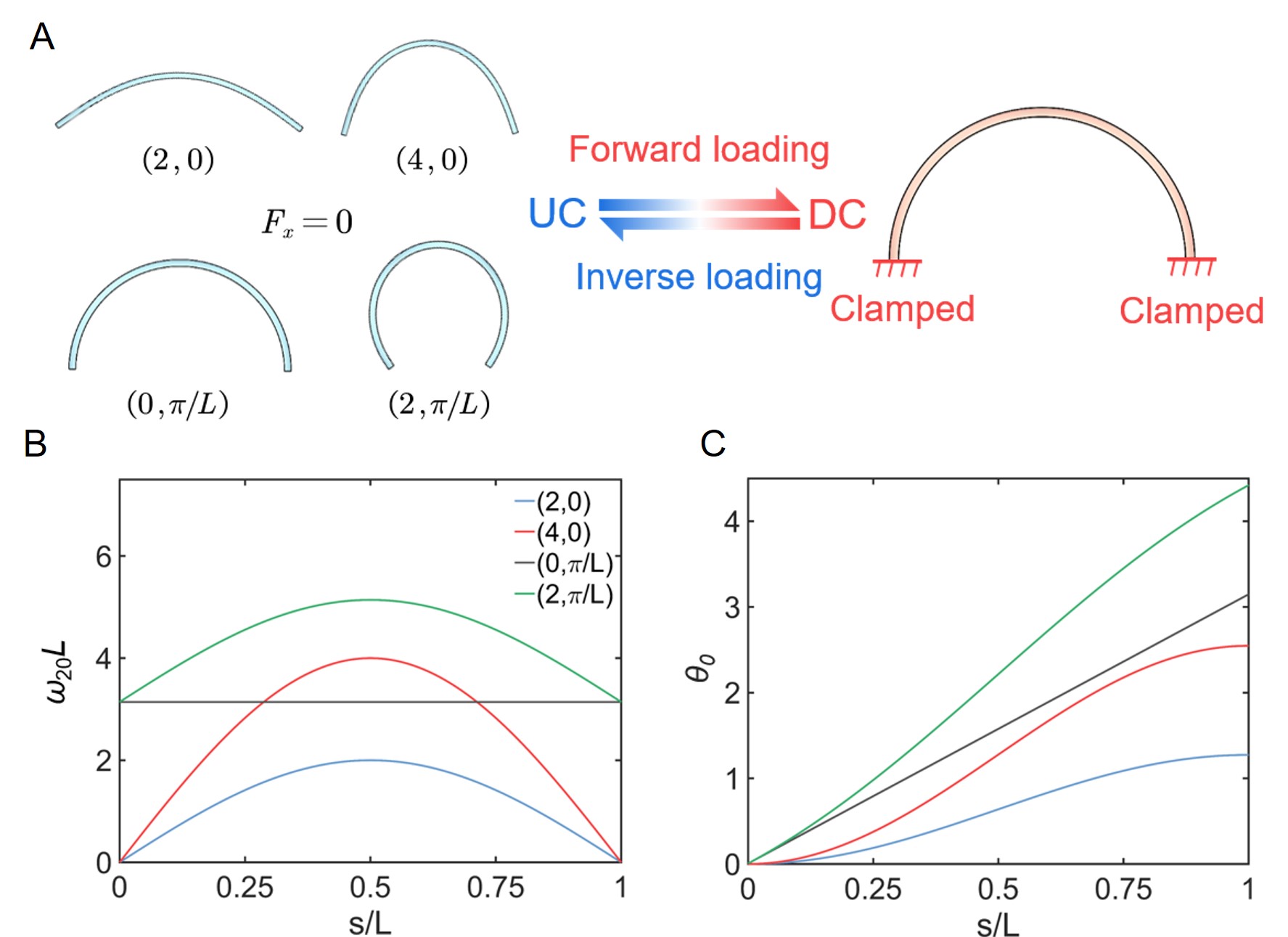}
    \caption{\textbf{The inverse design of an arc.} A family of UC deform into a target arc-shaped DC with displacement boundary condition applied. (A) UC deforming into the arc-shaped DC for differential parameters $(F_y/(\omega_2D_2),\omega_0)$. (B) Dimensionless material curvature as a function of normalized arc length $s/L$. (C) Rotation angle $\theta$ as a function of normalized arc length $s/L$ }
    \label{fig:f3}
\end{figure}

For the 2D case, we first expand $\mathbf{F}$ and $\mathbf{f}$ in local frame: $\{\mathbf{d}_3,\mathbf{d}_1\}$; $\mathbf{M}$ and $\mathbf{m}$ in local frame $\mathbf{d}_2$ and then use the motion equation of frame: $(\mathbf{d}_3)'=\omega_2\mathbf{d}_1$, $(\boldsymbol{d_1})'=-\omega_2\mathbf{d}_3$, where $\omega_2$ is the material curvature of DC, to simplify Eqs.~\eqref{eqn:1} $\sim$~\eqref{eqn:2}. It reads 

\begin{equation}
\begin{cases}
	\left( F_1 \right) '=-F_3\omega _2+f_1\\
	\left( F_3 \right) '=F_1\omega _2+f_3\\
	\left(M_2\right) '=-F_1+m_2\\
\end{cases},
    \label{eqn:17}
\end{equation}

where $F_i$ and $f_i$ is component of $\mathbf{F}$ and $\mathbf{F}$ on $\mathbf{d}_i$ ($i=1,3$),  $M_2$ and $m_2$ are components of $\mathbf{M}$ and $\mathbf{M}$ on $\mathbf{d}_2$. For the kirchhoff rod model with linear elastic material, the bending constitutive can be written as

\begin{equation}
    M_2=D_2(\omega_2-\omega_{20}).
    \label{eqn:18}
\end{equation}
The geometry of elastica meets the following relationship
\begin{equation}
	\left( \theta \right) '=\omega _2\\
,
    \label{eqn:19}
\end{equation}
where $\theta$ is the rotation angle of DC.
For the elastica without external load and moment $\bar{\mathbf{f}}$ and $\bar{\mathbf{m}}$, Eq.~\eqref{eqn:17} $\sim$~\eqref{eqn:19} can be simplified to the classic form derived by Lagrange in 1771~\citep{matsutani2024euler},
\begin{equation}
(D_2((\theta)'-\omega_{20}))'-F_x \sin\theta+F_y \cos\theta=0.
\label{eqn:20}
\end{equation}

For the forward problem, the goal is to solve for $\theta$, which renders Eq.~\eqref{eqn:20} nonlinear due to the simultaneous presence of its second derivative and trigonometric functions. When the UC is straight ($\omega_{20} = 0$), an analytical solution exists in terms of elliptic integrals, known as the \textit{elastica}~\citep{levien2008elastica,matsutani2010euler,matsutani2024euler}. For the inverse problem, $\theta$, which describes the DC is a known variable, and we need to solve $\omega_{20}$ instead of $\theta$.
Under this situation, the nonlinearity of Eq.~\eqref{eqn:20} is reduced, which is consistent with the first key point: reduced nonlinearity. As a result, $\omega_{20}$ can be directly obtained by integrating Eq.~(\ref{eqn:20}), i.e.,

\begin{equation}
\omega_{20} = \int_0^s \left( \theta_{ss} + \frac{F_y \cos \theta - F_x \sin \theta}{D_2} \right) ds + \omega_0.
\label{eqn:21}
\end{equation}
For Eq.~\eqref{eqn:21} the physical meaning of integration constant $\omega_0$ is that the clamped boundary can supply the arbitrary value of moment.
In analogy to the \textit{elastica}, we refer to the UC corresponding to a known DC as the \textit{inverse elastica}. The 2D inverse elastica is formally defined as:

\begin{definition}
A 2D inverse elastica corresponding to DC is a planar curve with material curvature $\omega_0$ given by

\begin{equation}
    \omega_{20} = \int_0^s \left( \theta_{ss} + \frac{F_y \cos \theta - F_x \sin \theta}{D_2} \right) ds + \omega_0,
    \label{eqn:22}
\end{equation}
where $D_2 \in (0, \infty]$, and $F_y$, $F_x$, and $\omega_0$ are real parameters.
\end{definition}

This definition raises a natural question: how should the parameters $F_y$, $F_x$, and $\omega_0$ be chosen, and what is their physical interpretation in the context of the inverse elastica?
To gain insight, we consider a toy model in which the DC is an arc, specified by $\theta = \omega_2 s$ and $\omega_2$ is constant. Substituting this expression into Eq.~\eqref{eqn:22} yields an analytical expression for the corresponding material curvature of UC $\omega_{20}$,

\begin{equation}
\omega_{20} = \frac{F_x \cos(\omega_2s) + F_y \sin(\omega_2s)}{\omega_2 D_2} + \omega_0.
\label{eqn:23}
\end{equation}

Eq.~(\ref{eqn:23}) provides an analytical expression for the material curvature $\omega_{20}$ corresponding to a given DC, determined by the external force components $F_x$, $F_y$, the bending stiffness $D_2$, and an integration constant $\omega_0$. To build physical intuition, consider a clamped–clamped beam deformed into a circular arc. We consider the pure bending state and let $F_x=0$, $F_y=0$ in Eq.~\eqref{eqn:23}, we find that $\omega_{20}=\omega_0$ is a constant. This result can be understood by considering two extreme examples: firstly, imagine that when we apply moments on the two ends of a straight beam, it will bend into an arc. This situation corresponds to $\omega_0=0$ and the UC is straight $\omega_{20}=0$. Secondly, if the UC is an arc with the same material curvature as DC, we neither need to apply the forces nor the moments on the two ends. This situation corresponds to $\omega_0=\omega_2$ and is consistent with Eq.~\eqref{eqn:23} for $F_x=F_y=0$.
While the inverse elastica in two aforementioned cases are consistent with intuitive expectations, a more rigorous investigation necessitates consideration of non-vanishing terminal loading conditions ($F_x \neq 0, F_y \neq 0$). Through the implementation of Eq.~\eqref{eqn:23}, non-trivial solutions emerge that exhibit characteristics beyond the predictive capacity of physical intuition (arc or straight line) alone as shown in Fig.~\ref{fig:f3}(A). The UCs are determined by the different parameters $(F_y/(\omega_2D_2),\omega_0)$. For example, the first UC $(2,0)$ means $F_y/(\omega_2D_2)=0, \ \omega_0=0$. The corresponding normalized curvature $\omega_{20}L$ and rotation angle $\theta_0$ of UC are illustrated in Figs.~\ref{fig:f3}(B) and (C), respectively.

Moreover, under clamped–clamped conditions, the parameters $F_x$, $F_y$, and $\omega_0$ are not uniquely determined by the shape $\theta(s)$. Since the inverse problem reconstructs $\omega_{20}$ from a known configuration, any combination of these parameters that satisfies Eq.~\eqref{eqn:23} is admissible. In other word, the same DC can result from a family of solutions parameterized by $(F_x, F_y, \omega_0)$, reflecting the second key point of inverse elastica theory mentioned in last section: the multiplicity of solutions of inverse problems arise from the non-uniqueness of $\boldsymbol{F}$ and $\boldsymbol{M}$. This inherent non-uniqueness is a fundamental feature of the inverse problem and must be considered when we want to select the desired UC. To understand the crucial concept of inverse elastica theory, the UC of a helix is obtained by applying inverse loading with numerical continuation in next section. 

\subsubsection{3D case: Inverse design of a helix}
\label{sec:helix}
We consider the DC as a helix, which means that both $\boldsymbol{\omega}$ and $\boldsymbol{\Omega}$ are constant. The component form of Eq.~\eqref{eqn:16} can be written as

\begin{equation}
\begin{cases}
	\left( F_1 \right)'=F_2\omega _3-F_3\omega _2+f_1\\
	\left( F_2 \right)'=F_3\omega _1-F_1\omega _3+f_2\\
	\left( F_3 \right)'=F_1\omega _2-F_2\omega _1+f_3\\
	D_1\left( \omega _1-\omega _{10} \right)'=D_2\left( \omega _2-\omega _{20} \right) \omega _3-D_3\left( \omega _3-\omega _{30} \right) \omega _2+F_2+m_1\\
	D_2\left( \omega _2-\omega _{20} \right)'=D_3\left( \omega _3-\omega _{30} \right) \omega _1-D_1\left( \omega _1-\omega _{10} \right) \omega _3-F_1+m_2\\
	D_3\left( \omega _3-\omega _{30} \right)'=D_1\left( \omega _1-\omega _{10} \right) \omega _2-D_2\left( \omega _2-\omega _{20} \right) \omega _1+m_3\\
	\left( q_{00} \right)'=-1/2\left( q_{01}\omega _{10}+q_{02}\omega _{20}+q_{03}\omega _{30} \right)\\
	\left( q_{01} \right)'=1/2\left( q_{00}\omega _{10}-q_{03}\omega _{20}+q_{02}\omega _{30} \right)\\
	\left( q_{02} \right)'=1/2\left( q_{03}\omega _{10}+q_{00}\omega _{20}-q_{01}\omega _{30} \right)\\
	\left( q_{03} \right)'=1/2\left( q_{01}\omega _{20}-q_{02}\omega _{10}+q_{00}\omega _{30} \right)\\
	\left( x_0 \right)'=2\left( q_{01}q_{03}+q_{00}q_{02} \right)\\
	\left( y_0 \right)'=2\left( q_{02}q_{03}-q_{00}q_{01} \right)\\
	\left( z_0 \right)'=2\left( q_{03}^{2}+q_{00}^{2}-1/2 \right)\\
\end{cases}.
\label{eqn:24}
\end{equation}
To simplify the control equations, we suppose there is no external load, which means $\bar{\mathbf{f}}=0$ and $\bar{\mathbf{m}}=0$. The analytical form of $\boldsymbol{\omega}_0$ is cumbersome, therefore we first consider the simplified case in the following part and use numerical continuation to solve Eq.~\eqref{eqn:24} directly. For the simplified case we suppose the UC lies on a plane and can only bend along the $\mathbf{d}_2$ direction, which means $\omega_{20}\neq0$, $\omega_{10}=0$, $\omega_{30}=0$.
The equation can be simplified as
\begin{equation}
    \begin{cases}
	F_1=\left( D_3-D_1 \right) \omega _1\omega _3\\
	F_2=\left( D_3-D_1 \right) \omega _2\omega _3\\
	F_3=\left( D_3-D_1 \right) \omega _{3}^{2}\\
	\omega _{20}=\left( 1-D_1/D_2 \right) \omega _2\\
\end{cases}.
\label{eqn:26}
\end{equation}

Eq.~\eqref{eqn:26} means that if we choose helix as DC and constrain the UC as a planar material curve, the solution is an unique arc, whose curvature is $(1-D_2/D_1)\omega_2$. For a circular cross section rod, $D_1=D_2$, which means the curvature of the UC is zero. This can be understood by the analytical solution of helix buckling from a straight rod~\citep{o2017modeling}.
Notice that Eq.~\eqref{eqn:24} can be solved by regulating the boundary displacement and rotation angle of UC, this is a similar process to solving the Kirchhoff rod equation under the displacement boundary condition. Hence, similar to the loading process in forward problem, here solving the inverse equation under the displacement boundary condition is defined as inverse loading. Note that we have already emphasized in Section~\ref{sec: Inverse problem} that inverse loading cannot be derived from the forward equation by simply swapping the geometry of UC and DC directly because the control equations of them are different. To understand this point we solve the forward loading and inverse loading as a comparison numerically. 

Numerical continuation is widely applied in solving the forward equation of Kirchhoff rod~\citep{doedel2007auto,yu2019bifurcations,yu2021numerical,yu2023continuous,shi2025double}. Inspired from this, we start from a helix spring with circular cross section as initial configuration and apply forward loading and inverse loading with numerical continuation, respectively. The equation of the helix is
\begin{equation}
  \begin{cases}
	x\left( t \right) =0.5\cos \left( 6\pi t \right)\\
	y\left( t \right) =0.5\sin \left( 6\pi t \right)\\
	z\left( t \right) =t\\
\end{cases}\,\,  t\in \left[ 0.1,1.1 \right] .
\label{eqn:27}
\end{equation}

\begin{figure}[ht]
    \centering
    \includegraphics[width=1.0\columnwidth]{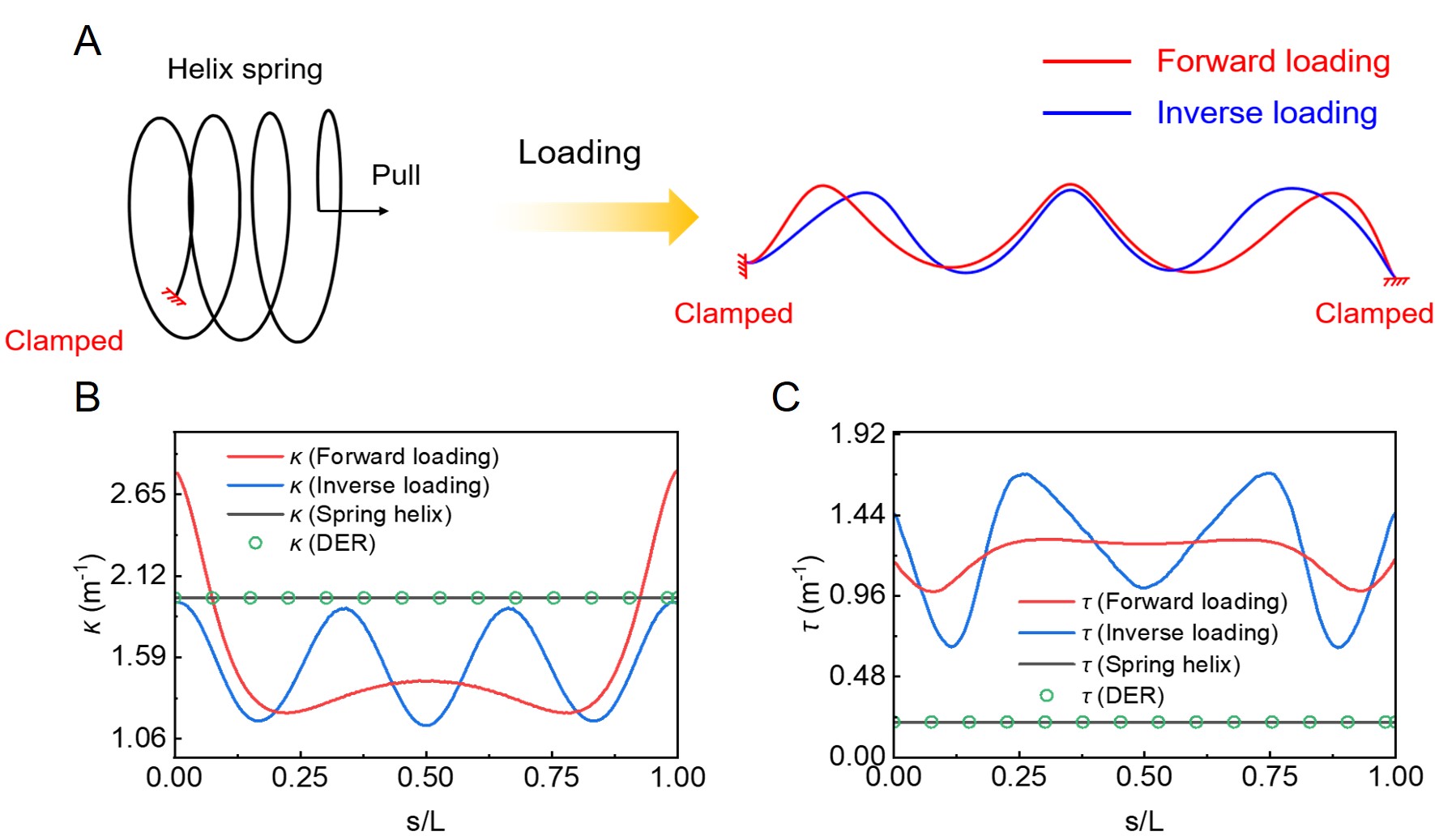}
    \caption{\textbf{Comparison between inverse loading and forward loading.}
    (A) Deformation of a helix spring under inverse and forward loading. (B) Curvature distribution of the helix after inverse loading and forward loading respectively. (C) Torsion distribution of the helix after inverse and forward loading. The green points represent numerical results from DER verification.}
    \label{fig:f4}
\end{figure}

In the inverse loading, we control the z coordination of the helix spring by stretching the right end from 1.1 to 6.1 as shown in Fig.~\ref{fig:f4}(A). During the process of forward loading, the radius of the middle part of the spiral spring decreases (the red spring). However, for the inverse loading, the variation of the radius of the middle part is not apparent as shown in Fig.~\ref{fig:f4}(A) (the blue spring). This phenomenon is easily understandable based on common sense. When a helical spring is stretched, the radius of the helix near the midpoint tends to decrease noticeably. In contrast, compressing the spring by an equivalent amount results in a much less pronounced change in the radius at the center. Indeed, as shown in Figs.~\ref{fig:f4}(B) and (C), the changes of curvature and torsion from 0.25 to 0.75 are relatively gradual for the red line, which means the middle part in the forward loading tends to unwind. But for the UC obtained from inverse loading, both of the curvature and torsion change relatively drastically. The difference between forward loading and inverse loading originates from the difference between the elastica theory and the inverse elastica theory as discussed in Section~\ref{sec: Inverse problem}. We also use DER to verify our inverse theory by compressing the UC obtained from inverse loading, the curvature and torsion of DC is consistent with the initial configuration as shown in Figs.~\ref{fig:f4}(B) and (C).

In this section, two illustrative cases—namely, the inverse design of an arc and inverse loading of a helix spring, have been investigated to understand the three key points of inverse elastica theory in Section~\ref{sec: Inverse problem}. Although we have given the method to solve the UC by inverse loading, a predefined loading path is required to carry out inverse loading on the DC. Sometimes, we want the UC to exhibit certain characteristics that are difficult to be expressed as explicit boundary conditions. For example, we want to minimize the height variance of the helix discretized surfaces to reduce the volume. Under this situation, solving the inverse elastica equation as a BVP is generally not an effective approach for inverse design. To address this challenge, an theory-assisted optimization strategy is proposed. In the next section, several examples such as knots and helix discretized torus, cones, spheres, and hyperboloids are investigated using optimization approach. And finally, by using the optimization approach, we achieve the inverse design of the conformable hemispherical helix ribbon with minimal volume, which is usually used in the fabrication of deployable and conformable hemispherical antennas.

\section{Inverse design enabled by theory-assisted optimization}

In Section~\ref{sec: framework}, we have developed a general inverse design theory and understood the three characteristics underlaying the inverse elastica theory by solving the inverse equation as BVP analytically and numerically. In fact, under some situations our goal is to select the UC with certain characteristics. However, the certain characteristics that we want for the UC are not easy to be expressed as the boundary condition for Eq.~\eqref{eqn:16}. For example, if we want UC to be as close as possible to a straight line, which means the curvature of UC should be as small as possible, the problem can be summarized as finding the initial value to minimize the curvature: $\min_{\left\{ \boldsymbol{F}\left( 0 \right) ,\boldsymbol{M}\left( 0 \right) \right\}} \int_0^L{\omega_{10}^{2}+\omega_{20}^{2}}ds$, where $\omega_{10}$ and $\omega_{20}$ are the material curve of UC in width and thickness directions. This is an optimization problem and it is not easy to express this condition as boundary condition for Eq.~\eqref{eqn:16}. Therefore, we propose a theory-assisted optimization method to achieve some complex inverse design cases.

\begin{figure}[b!]
    \centering
    \includegraphics[width=0.75\columnwidth]{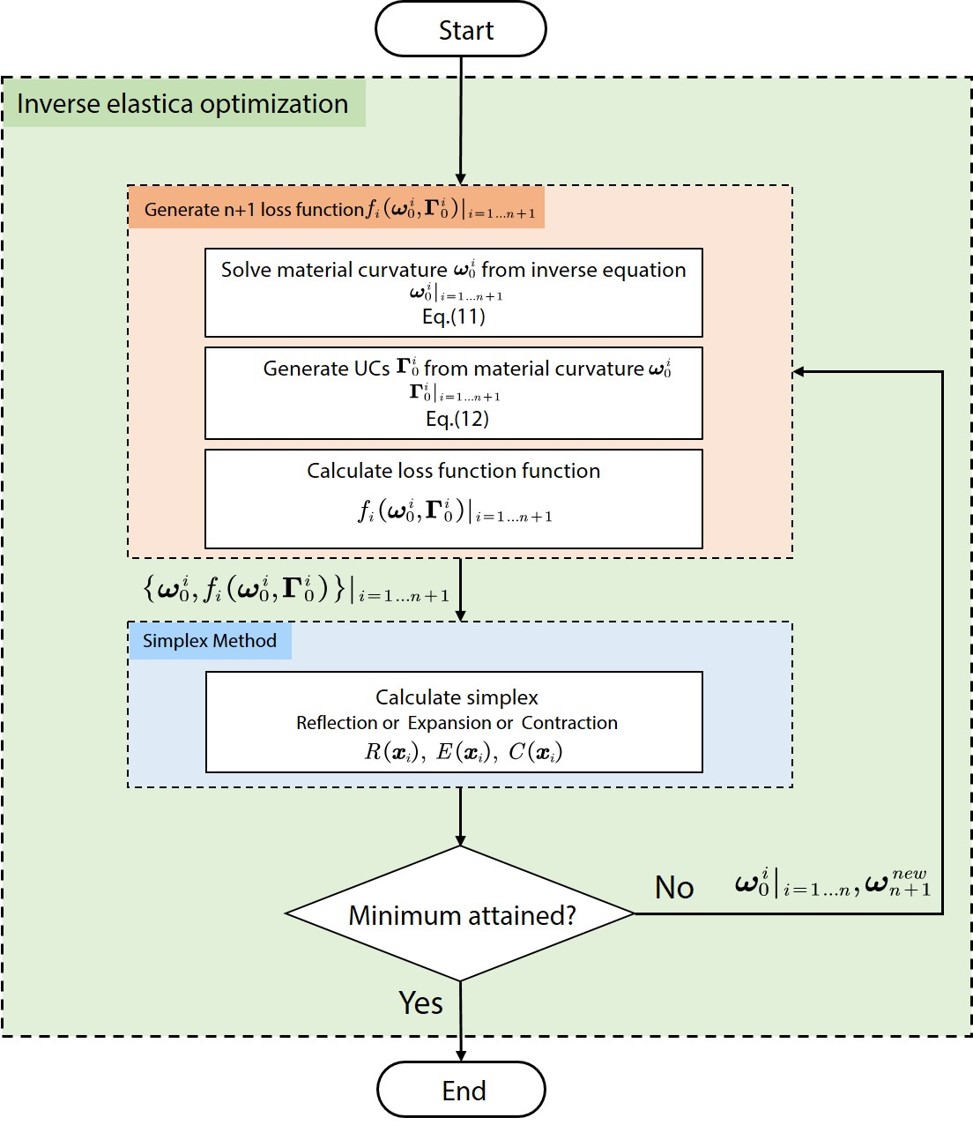}
    \caption{\textbf{The algorithm flow chart of theory-assisted optimization.}}
    \label{fig:f5}
\end{figure}

\subsection{Theory-assisted optimization}
\label{sec:opt}
As shown in  Fig.~\ref{fig:f5}, we first define the inverse design as an optimization problem:
\begin{equation}
 \begin{aligned}
    \left\{ \bar{\mathbf{F}}^*\left( 0 \right) , \bar{\mathbf{M}}^*\left( 0 \right) \right\} =\underset{\left\{ \boldsymbol{F}\left( 0 \right) ,\boldsymbol{M}\left( 0 \right) \right\}}{\min}f\left( \boldsymbol{\omega }_0, \boldsymbol{\Gamma}_0 \right),  \\
s.t. \mathcal{I} nv\mathcal{E} elastica\left( \boldsymbol{\omega }_0 \right) =\boldsymbol{0},
\end{aligned}
\label{eqn:28}
\end{equation}
where $\left\{ \bar{\mathbf{F}}^*\left( 0 \right) , \bar{\mathbf{M}}^*\left( 0 \right) \right\}\in \mathcal{R}^{6\times1}$ is the optimal initial value of force and moment, $\mathcal{I} nv\mathcal{E} lastica\left( \boldsymbol{\omega }_0 \right)$ describes the underlying physics constraints from the inverse elastica theory in Eq.~\eqref{eqn:16}, and $f(\boldsymbol{\omega}_0,\boldsymbol{\Gamma}_0)$ is the loss function, which represents the certain characteristics we want UC to possess. Here we combine inverse elastica theory and Nelder-Meaed method~\citep{singer2009nelder} to achieve the optimization. As shown in Fig.~\ref{fig:f5}, fistly we give an initial guess of $\left\{ \bar{\mathbf{F}}\left( 0 \right) ,\bar{\mathbf{M}}\left( 0 \right) \right\}$, then the material curvature $\boldsymbol{\omega}_0$ is calculated from inverse elastica theory Eq.~\eqref{eqn:16}. Using the geometry equation, the coordination of UC $\boldsymbol{\Gamma}_0$ can be calculated from the material curvature $\boldsymbol{\omega}_0$. Finally $\{\boldsymbol{\omega}_0,f(\boldsymbol{\omega_0},\boldsymbol{\Gamma}_0)\}$ is obtained as the input for Nelder-Mead method. The Nelder-Mead algorithm is a direct search method that optimizes the loss function f(x) without requiring gradient information. Starting from an initial simplex of 7 points $\boldsymbol{\omega}_0^i (i=1...7)$ in six-dimensional space, it iteratively updates the simplex through reflection, expansion, contraction, and shrink operations based on the function values $f(\boldsymbol{\omega}_0^i,\boldsymbol{\Gamma}_0)$ at each vertex. The algorithm replaces the worst vertex with a better point in each iteration, gradually moving the simplex toward the minimum region. The process continues until the simplex size falls below a specified tolerance or the function value improvement becomes negligible, providing an approximate solution for the minimization of $f(\boldsymbol{\omega_0},\boldsymbol{\Gamma}_0)$. Some demonstrations are presented with the proposed theory-assisted optimization strategy in Section~\ref{sec:examples}.

\subsection{Inverse design demonstration}
\label{sec:examples}
\subsubsection{The trefoil knot}

\begin{figure}[ht]
    \centering
    \includegraphics[width=1.0\columnwidth]{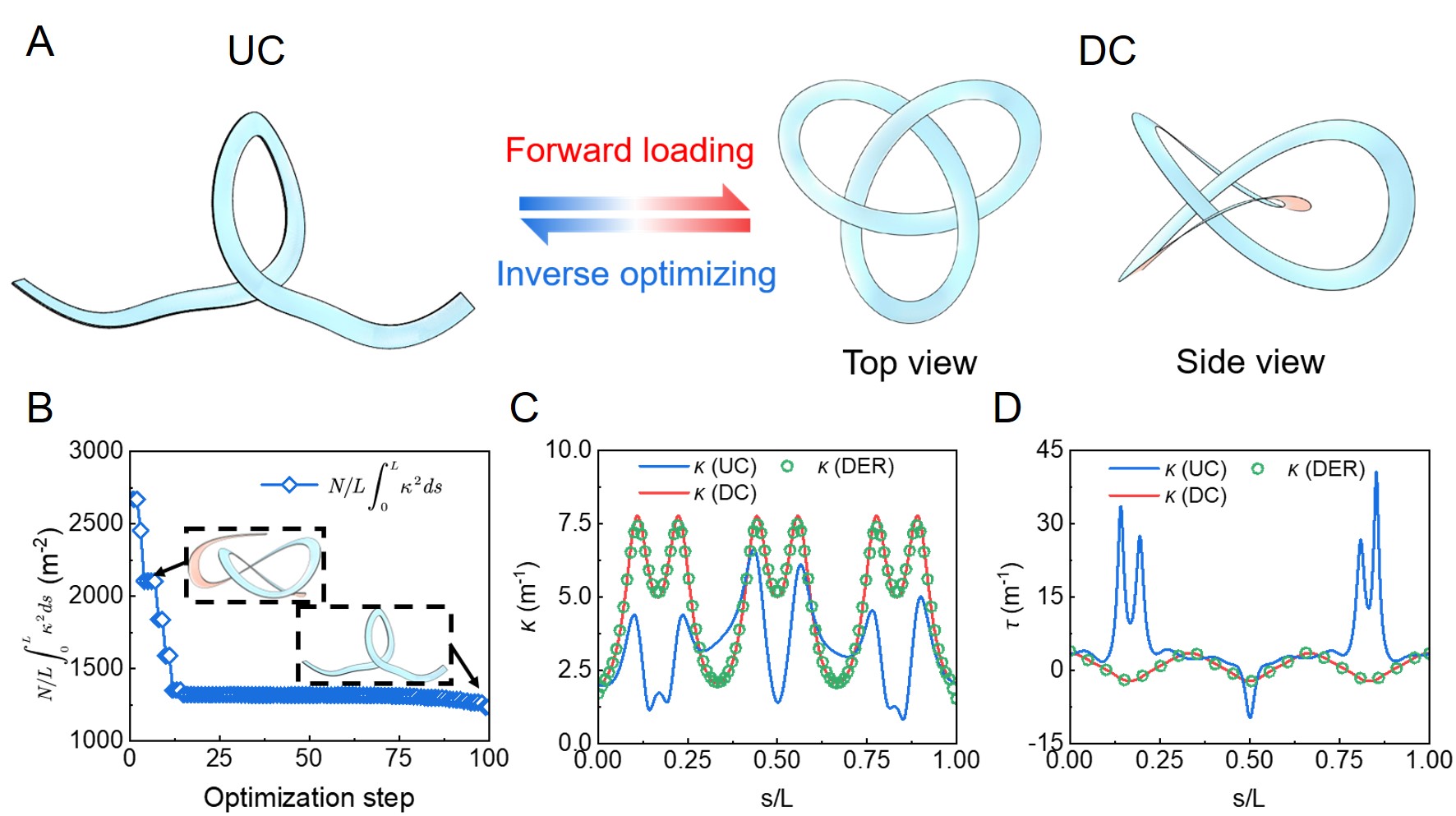}
    \caption{\textbf{Inverse design of a trefoil knot.} (A) Elastic deformation from an open ribbon as UC to a trefoil knot as DC. (B) The value of the loss function during the optimization process. (C) The curvature of UC (blue line), DC (red line) and DER verification (green points). (D) The torsion of UC (blue line), DC (red line) and DER verification (green points).}
    \label{fig:f6}
\end{figure}

In previous study, \cite{moulton2018stable} demonstrated that a ribbon can deform to a stable open knot, but a rod with circular cross section cannot. Inspired from this work, we try to find a ribbon as UC with the smallest possible curvature to deform into a close trefoil knot. The close trefoil knot means that a ribbon whose middle line is along with the standard trefoil knots in Eq.~\eqref{eqn:29} and the direction of the normal vector is continuous; hence the twisting angle between the reference frame and material frame is set as $(\pi-0.9192)*t$ to ensure the direction of normal vectors at the two ends are the same as shown in Fig.~\ref{fig:f6}(A).

\begin{equation}
    \begin{cases}
	x\left( t \right) =\sin \left( 2\pi t \right) +2\sin \left( 4\pi t \right)\\
	y\left( t \right) =\cos \left( 2\pi t \right) -2\cos \left( 4\pi t \right)\\
	z\left( t \right) =-\sin\mathrm{(}6\pi t)\\
\end{cases}\,\,  t\in \left[ 0,1.0 \right). 
\label{eqn:29}
\end{equation}
The loss function is defined as:
\begin{equation}
    f\left( \boldsymbol{\omega }_0,\mathbf{\Gamma }_0 \right) =\underset{\left\{ \boldsymbol{F}\left( 0 \right) ,\boldsymbol{M}\left( 0 \right) \right\}}{\min}\frac{N}{L}\int_0^L{\left( \omega _{10}^{2}+\omega _{20}^{2} \right)}ds,
    \label{eqn:30}
\end{equation}
where $N$ is the number of the discrete point.
By minimizing Eq.~\eqref{eqn:30}, the optimized UC is obtained, as shown in Fig.~\ref{fig:f6}(A). The evolution of the loss function during the optimization process is illustrated in Fig.~\ref{fig:f6}(B). Notably, a sharp decrease from approximately 2100 to 1300 around the 10th iteration suggests a sudden transition between two quasi-steady states. The curvature and torsion profiles of the DC and the UC are presented in Figs.~\ref{fig:f6}(C) and (D), respectively. It is observed that the optimized UC maintains geometric symmetry with respect to the midpoint, i.e., at $s/L = 0.5$. To verify our theory, DER simulation is performed and the results are shown with the green points in Figs.~\ref{fig:f6}(C) and (D). The simulation results are consistent with our theoretical solution and the forward loading video is shown in~\ref{app:C}.

In this section, we find the UC with minimal curvature by our inverse elastica theory for a close trefoil knot as DC. In the following section, we are more concerned about finding the UC of helix discretized curved surfaces with different Gaussian curvatures.

\subsubsection{The torus}
The inverse elastica theory enables the design of UC for spatial curves exhibiting complex topological features. Moreover, it facilitates the construction of UCs for specific types of DC, such as those that helix discretized curved surfaces with various Gaussian curvatures.
\begin{figure}[ht]
    \centering
    \includegraphics[width=1.0\columnwidth]{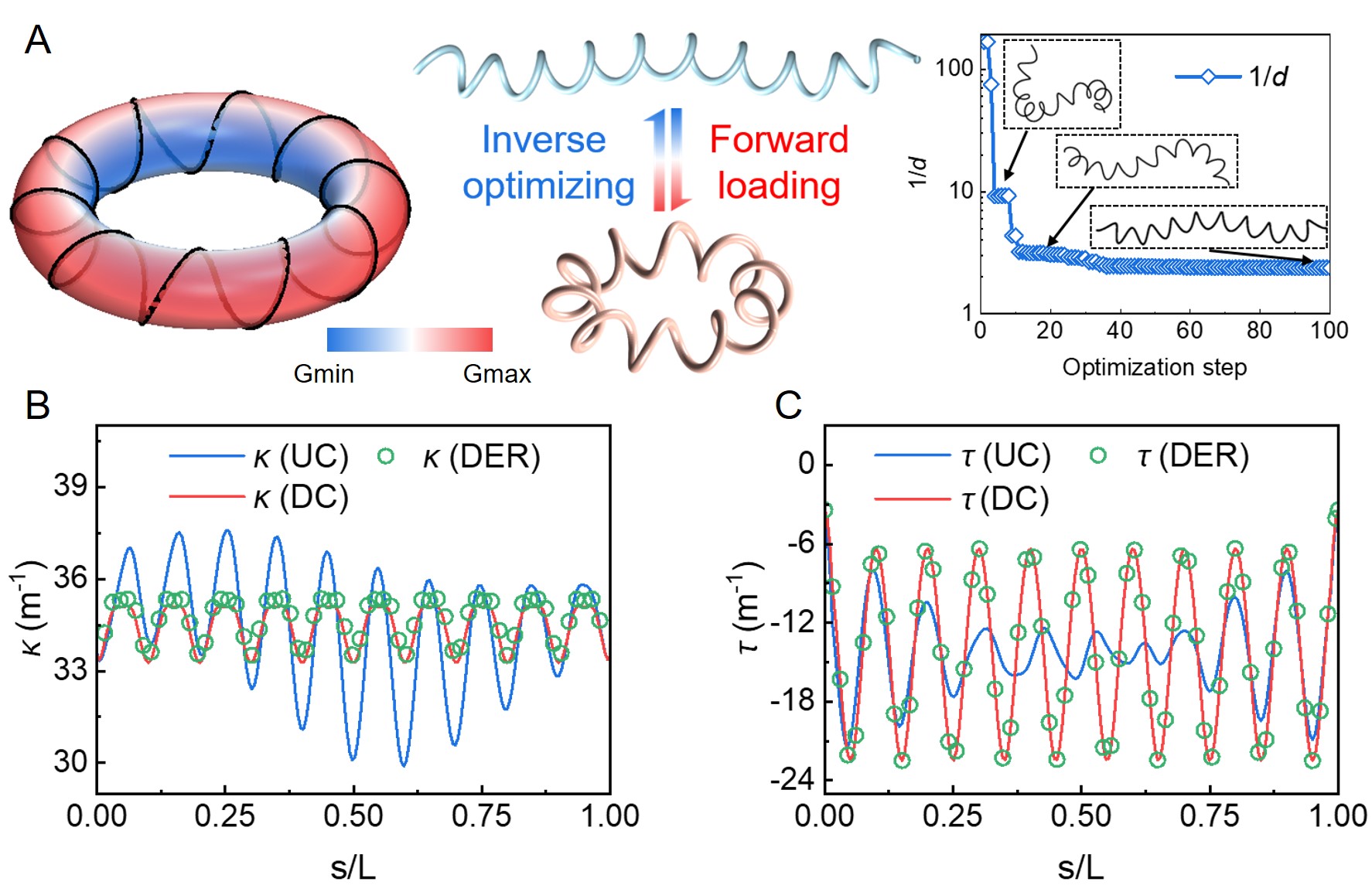}
    \caption{\textbf{Inverse design of a helix discretized torus.} A helix discretized torus exhibiting both negative and positive Gaussian curvature, with the colormap ranging from blue (minimum Gaussian curvature) to red (maximum Gaussian curvature) (B) Elastic deformation from a spring as UC to a helix discretized torus as DC. (C) The value of the loss function during the optimization process. (D) The curvature of UC (blue line), DC (red line) and DER verification (green points). (E) The torsion of UC (blue line), DC (red line) and DER verification (green points).}
    \label{fig:f7}
\end{figure}
The torus is a geometrically rich surface characterized by regions of both positive and negative Gaussian curvature. Such configurations are frequently encountered in various biological systems in nature~\citep{wang2023curvature,wang2025nonlinear}.
Here we choose a helix discretized torus as the second example as shown in Fig.~\ref{fig:f7}(A).  The equation of the helix discretized torus can be written as:
\begin{equation}
    \begin{cases}
        x(t)=(r+R\cos(2\pi N t))\cos(2\pi t)\\
        y(t)=(r+R\cos(2\pi N t))\sin(2\pi t)\\
        z(t)=r\sin(2\pi Nt)
    \end{cases},
\label{eqn:31}
\end{equation}
where $R$ and $r$ are the radius of torus respectively, $N$ is a positive integer and determines the number of turns of the spiral wire on the torus. For this case, we choose $R=4r=0.1$, $N=10$. Here, the UC is desired to be as slender as possible to facilitate compact storage. Accordingly, the objective function is formulated as:
\begin{equation}
    f(\boldsymbol{\omega}_0,\boldsymbol{\Gamma}_0)=1/d=\frac{1}{\sqrt{(x(0)-x(1))^2+(y(0)-y(1))^2+(z(0)-z(1))^2}}.
\end{equation}

For the torus, its volume can be approximated by that of the circumscribed cylinder, given by $2\pi r(R + r)^2$, while the volume of the slender UC is estimated as $2\pi^2 r^2 R$. Accordingly, the volume ratio between the UC and the DC is approximated as $\frac{\pi r / R}{(1 + r / R)^2}$.
In this study, we set $r/R = 0.25$, resulting in a volume reduction ratio of approximately 0.5, indicating that the volume of the UC is only half that of the DC.
As shown in Fig.~\ref{fig:f7}(B), the optimized UC exhibits symmetry but deviates from a conventional straight helix, suggesting that a standard toroidal helix (Eq.~\eqref{eqn:31}) cannot be obtained by simply clamping the ends of a straight helical spring. In Fig.~\ref{fig:f7}(C), three plateau regions are observed in the optimization curve, indicating the occurrence of steady-state transitions during the optimization process, similar to those encountered in the forward problem~\citep{yu2019bifurcations,yu2021numerical,yu2023continuous}.
Although a detailed bifurcation analysis of the inverse elastica equations could be performed using continuation methods~\citep{doedel2007auto,yu2019bifurcations}, our objective here is to obtain UCs with specific geometric characteristics. Therefore, instead of analyzing the bifurcation process in detail, we adopt a direct optimization approach to obtain the desired UC. The curvature and torsion distributions of the DC and UC are shown in Figs.~\ref{fig:f7}(D) and (E), respectively. The accuracy of the inverse elastica solution is further validated through forward simulation using the discrete elastic rod (DER) method, demonstrating the correctness of the proposed inverse design theoretical framework.

\subsubsection{Curve discretized surfaces}

Previously, we investigated the toroidal helix as a discretization of a curved surface exhibiting both positive and negative Gaussian curvatures simultaneously. In this section, we extend the study to three representative helix discretized curved surfaces: the cone, the sphere, and the hyperboloid, which typify surfaces with zero, positive, and negative Gaussian curvature, respectively.

\begin{figure}[ht]
    \centering
    \includegraphics[width=1.0\columnwidth]{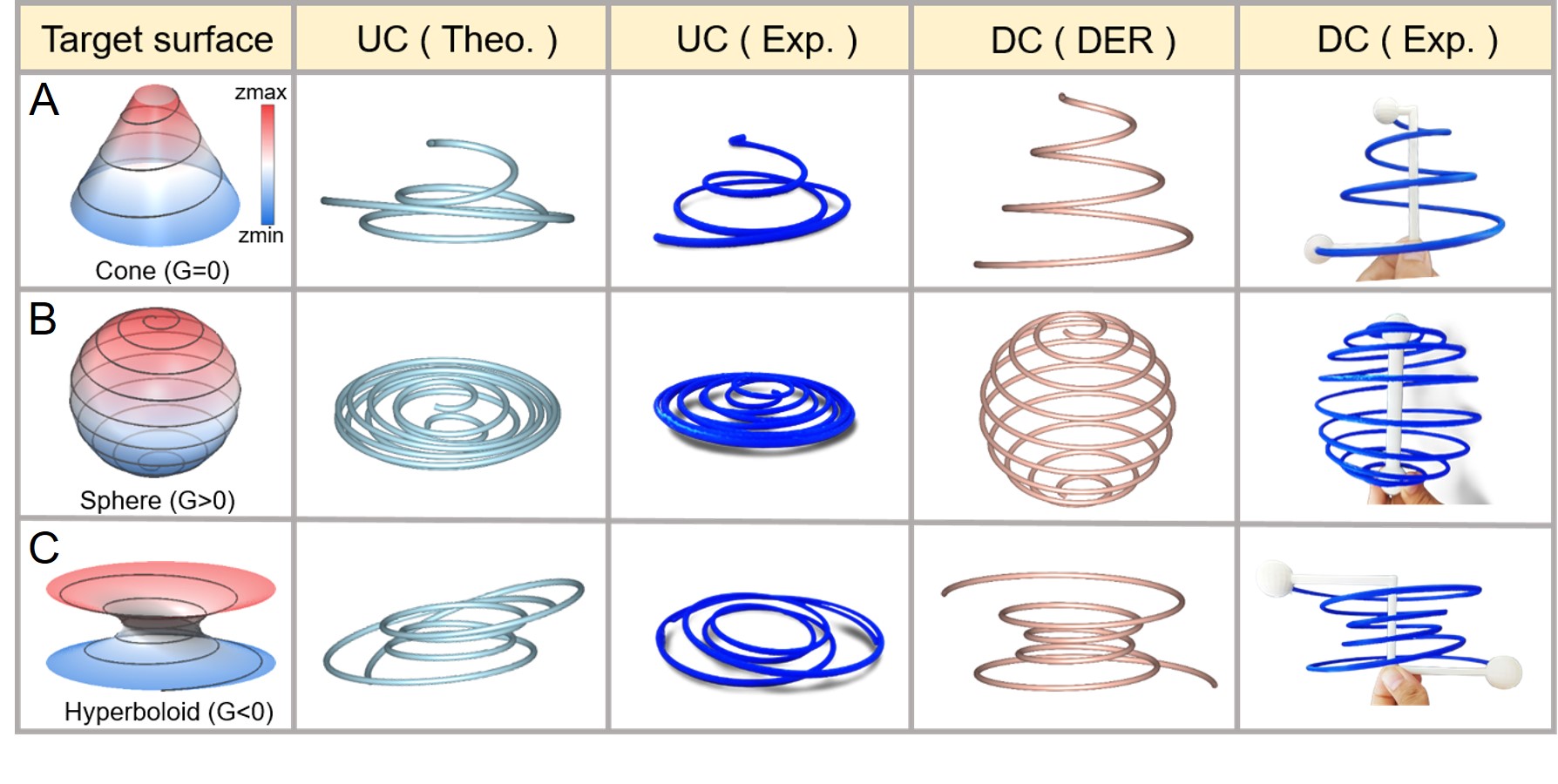}
    \caption{\textbf{Inverse design of three helix discretized surfaces with zero (cone), positive (sphere) and negative (hyperboloid) Gaussian curvatures.} (A) Inverse design of a helix discretized cone with zero Gaussian curvature. (B) Inverse design of a helix discretized sphere with positive Gaussian curvature. (C) Inverse design of a helix discretized hyperboloid with negative Gaussian curvature.}
    \label{fig:f8}
\end{figure}

For these three surfaces, the specific equations are shown in~\ref{app:A} and our goal is to minimize the height variance in order to reduce spatial volume occupancy. Accordingly, we define the loss function as the variance of the $z$-coordinate, which serves as a measure of vertical compression. This can also be interpreted as:
\begin{equation}
    f(\boldsymbol{\omega}_0,\boldsymbol{\Gamma}_0)=\frac{1}{L}\int_0^L(z-\int_0^Lzds/L)^2ds.
\end{equation}
As shown in Fig.~\ref{fig:f8}, the UC obtained by our theory is well consistent with the experiments and DER simulations. The details of the experiment settings is shown in~\ref{app:A}.

\begin{figure}[ht]
    \centering
    \includegraphics[width=1.0\columnwidth]{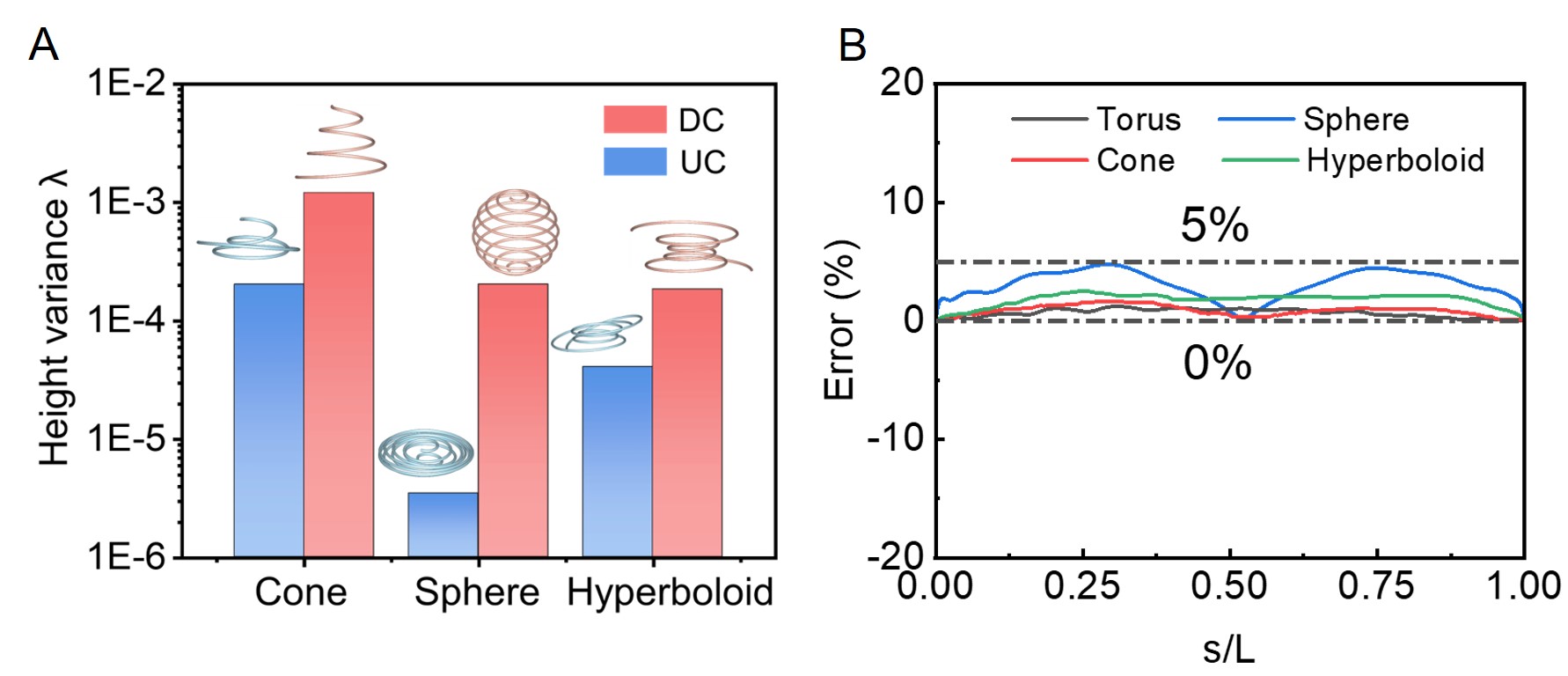}
    \caption{\textbf{The height variance of helix discretized surfaces with different Gaussian curvatures and the error analysis between theoretical results and DER verifications} (A) Height variance of DC and UC for helix discretized cone, sphere and hyperboloid surfaces respectively. (B) The error between target shape and DER verification results for helix discretized torus, cone, sphere and hyperboloid surfaces respectively. }
    \label{fig:f9}
\end{figure}

The optimized height variance is illustrated in Fig.~\ref{fig:f9}(A). The results demonstrate that a spherical helix can be achieved by stretching a very flat UC. Notably, previous studies have required complex boundary conditions—such as impact or multi-point compression—to morph a structure into a spherical shape~\citep{fan2020inverse,lee2020computational,liu2020tapered,kansara2023inverse}. In contrast, our approach enables the design of such UCs using inverse elastica theory, allowing the structure to transform into a helix-discretized sphere simply by clamping the two ends.
As shown in Fig.~\ref{fig:f9}(A), the height variance of cone is reduced from $1.2e^{-2}$ to $2.0e^{-4}$, the height variance of sphere is reduced from $2.0e^{-4}$ to $3.5e^{-6}$ and the height variance of hyperboloid is reduced from $1.8e^{-4}$ to $4.0e^{-5}$, respectively. The height variance of these three types of helix discretized surfaces are reduced by one to two orders of magnitude. 

It is worth noting that the height variance of the cone and hyperboloid helices are not as small as that of the sphere. This behavior can be interpreted through the concept of inverse loading, as introduced in the Section~\ref{sec:helix}.
Inverse loading, though analogous in outcome to forward loading, arises from a fundamentally different physical mechanism. Viewing the optimization process as a form of inverse loading provides insight into the results. When a cone or hyperboloid helix is compressed under clamped–clamped boundary conditions, the central region tends to bulge outward (as seen in Figs.~\ref{fig:f8}(A) and (C)). This geometric response increases the height variance under large compression, thereby limiting the achievable height variance.

All of the cases are verified with DER simulations, the calculated error between our theory and DER verifications as shown in Fig.~\ref{fig:f9}(B), the error is calculated by Eq.~\eqref{eqn:34}:
\begin{equation}
    \mathrm{Error}=\sqrt{(x_0^i-x_{DER}^i)^2+(y_0^i-y_{DER}^i)^2+(z_0^i-z_{DER}^i)^2}/L,
    \label{eqn:34}
\end{equation}
where $x_0^i, y_0^i, z_0^i$ are the discretized points of DC, $x_{DER}^i, y_{DER}^i, z_{DER}^i$ are the verified results from DER.
The numerical results show that the error of our theory is controlled within 5 \% as shown in Fig.~\ref{fig:f9}(B). 

In summary, this section demonstrates the inverse design of helix discretized curved surfaces using the inverse elastica framework. To further illustrate the practical relevance of our inverse elastica theory, an engineering application is presented in Section~\ref{sec:engineering}.

\subsection{Engineering application: inverse design of a conformable hemispherical helix antenna}
\label{sec:engineering}
In Section~\ref{sec:examples}, we demonstrated the inverse design of curved surfaces discretized using a complex helical curves. However, in certain engineering applications, it is necessary to design a ribbon that conforms to a target surface. Therefore, in this section, we extend the framework to the inverse design of a conformable ribbon, where conformable implies that the ribbon’s normal direction aligns with the normal vector of the target curved surface. To achieve this, we first establish the geometric foundation for conformable ribbons on curved surfaces. 

In our analysis, the conformable ribbon is modeled as a ruled surface:
\begin{equation}
    \boldsymbol{r}(s,u)=\boldsymbol{\Gamma}(s)+u\boldsymbol{v}.
\end{equation}
Here, $s$ denotes the arc-length parameter along the ribbon’s centre line, and $u$ is a transverse parameter associated with the ribbon’s width. The vector $\boldsymbol{v} = a\mathbf{d}_3 + b\mathbf{d}_1$, with $\sqrt{a^2 + b^2} = 1$, is a unit vector lying in the $\{\mathbf{d}_3, \mathbf{d}_1\}$ plane, where $\mathbf{d}_3$ is the tangent vector and $\mathbf{d}_1$ is the material director along the transverse direction of the ribbon.
For a conformable ribbon, the surface normal of the ribbon coincides with the normal vector of the target curved surface. As a result, the material frame of the ribbon is aligned with the Darboux frame of the surface~\citep{do2016differential,abbena2017modern}.
The material frame of the ribbon can be written as:
\begin{equation}
     \left(\begin{array}{c}
	\mathbf{d}_3'\\
	\mathbf{d}_1'\\
	\mathbf{d}_2'\\
\end{array} \right)=\left( \begin{matrix}
	0&		\kappa _g&		\kappa _n\\
	-\kappa _g&		0&		\tau _g\\
	-\kappa _n&		-\tau _g&		0\\
\end{matrix} \right) \left( \begin{array}{c}
	\mathbf{d}_3\\
	\mathbf{d}_1\\
	\mathbf{d}_2\\
\end{array} \right), 
\end{equation}
where $\kappa_n$ is the normal curvature, $\kappa_g$ is the geodesic curvature and $\tau_g$ is the geodesic torsion.
For a conformable ribbon, since its normal direction is aligned with that of the target curved surface, the tangent plane of the ribbon must coincide with the tangent plane of the surface at every point. Consequently, the $\mathbf{d}_2$ component of the partial derivative $\partial_s \boldsymbol{r}(s, u)$ must vanish, i.e., $\mathbf{d}_2 \cdot \partial_s \boldsymbol{r}(s, u) = 0$, where $\mathbf{d}_2$ denotes the surface normal direction in the Darboux frame.
\begin{equation}
    \boldsymbol{r}'(s,u)=u\left( a\kappa _n+b\tau _g \right) \mathbf{d}_2+\left( 1-ub\kappa _g \right) \mathbf{d}_3+ua\kappa _g\mathbf{d}_1.
\end{equation}
Thus, we have $a\kappa_n+b\tau_g=0$, and finally the equation of the ribbon can be derived as:
\begin{equation}
    \boldsymbol{r}(s,u)=\boldsymbol{r}(s)+\frac{u}{\sqrt{\kappa_n^2+\tau_g^2}}(\tau_g\mathbf{d}_3-\kappa_n\mathbf{d}_1).
    \label{eqn:38}
\end{equation}
Using Eq.~\eqref{eqn:38}, we can derive conformable ribbon with the given target surface and the helix space curve.

\begin{figure}[ht]
    \centering
    \includegraphics[width=1.0\columnwidth]{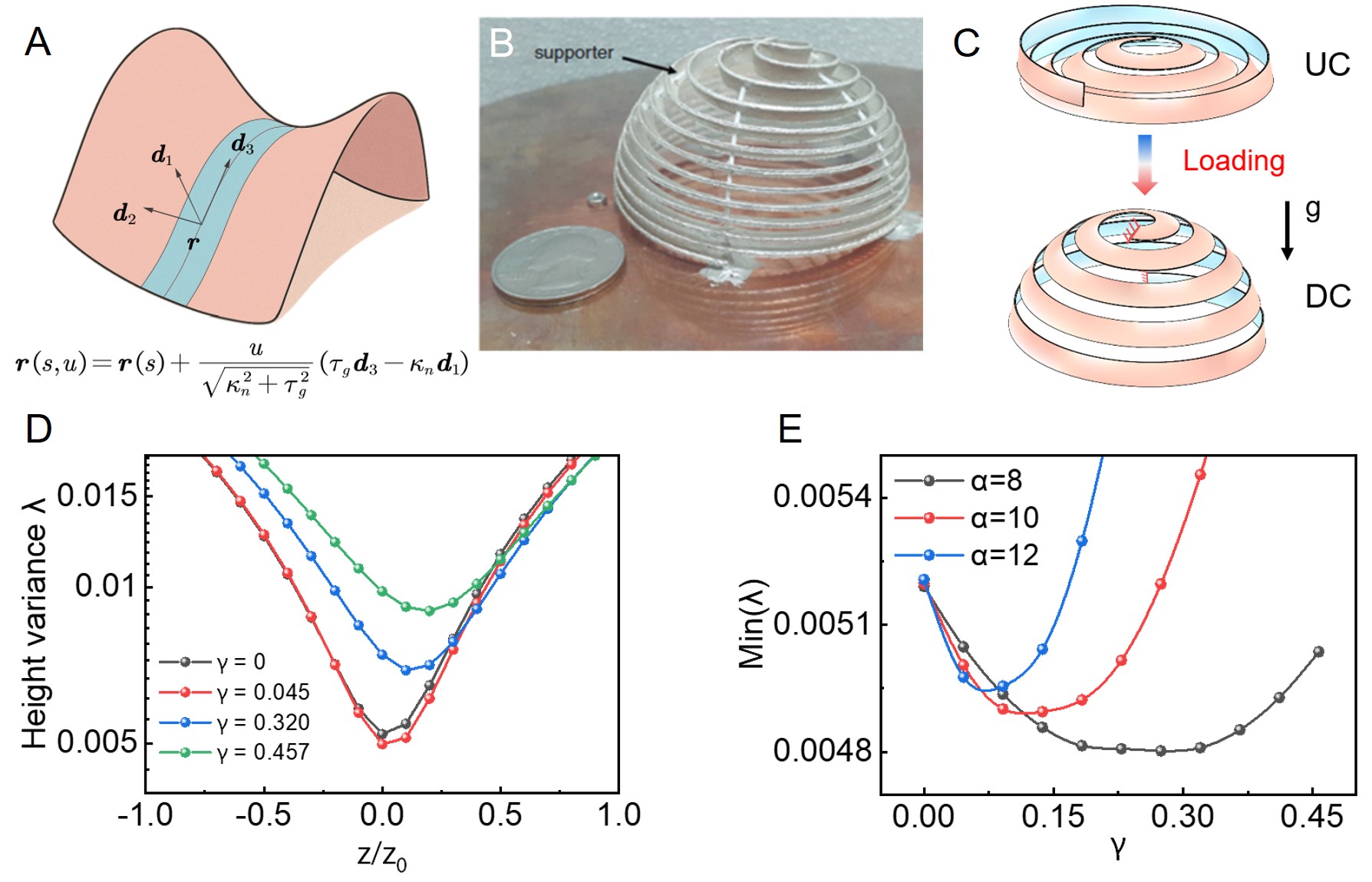}
    \caption{\textbf{The optimal inverse design of a  deployable and conformable hemispherical helix small antenna.} (A) The illustration of a deployable and conformable ribbon. (B) The small hemispherical helix antennas with supporter~\citep{kong2016electrically}. (C) Morphing of conformable hemispherical helix ribbon by clamped-clamped boundary condition with gravity effect considered. (D) The height variance with different elastic-gravity parameter as a function of normalized height $z/R$. (E) The minimal value of height variance $\lambda$ for different width-thickness ratio $\alpha$ as a function of elastic-gravity parameter $\gamma$.}
    \label{fig:f10}
\end{figure}

The hemispherical helix is a critical morphology for antenna systems due to its ability to approach the Chu–Harrington limit~\citep{chu1948physical,kong2016electrically}, making it highly valuable in radio frequency applications. Its unique geometry enhances radiation efficiency while maintaining a small form factor, which is essential for modern miniaturized radar designs. As shown in Fig.~\ref{fig:f10}(B), for the traditional hemispherical helix antennas the supporter is necessary to overcome the effect of gravity, but we can achieve the inverse design of deployable and conformable hemisphere helix ribbon with only two clamped ends by inverse elastica theory. Here, we consider the inverse design of a conformable hemispherical helix (Eq.~\eqref{eqn:39}).
\begin{equation}
    \begin{cases}
	x\left( t \right) =R\sin \left( \pi \left( 2-t \right) /4 \right) \cos \left( 5\pi \left( 2-t \right) \right)\\
	y\left( t \right) =R\sin \left( \pi \left( 2-t \right) /4 \right) \sin \left( 5\pi \left( 2-t \right) \right)\\
	z\left( t \right) =R\cos \left( \pi \left( 2-t \right) /4 \right)\\
\end{cases}\,\,  t\in \left[ 0.1,1.9 \right]. 
\label{eqn:39}
\end{equation}

For ease of transportation, our objective is to design a conformable hemispherical helical ribbon with minimal height variance, as illustrated in Fig.~\ref{fig:f10}(C). Two key dimensionless physical parameters are considered in the design: the gravity–elasticity number, defined as
$\gamma = \frac{\rho g L^3}{E w^2}$, and the width-to-thickness ratio, $\alpha = \frac{w}{t}$, where $w$ and $t$ denote the ribbon's width and thickness, respectively, $\rho$ is the material density, $g$ is the gravitational acceleration, $L$ is the arc length of the ribbon, and $E$ is the Young’s modulus. Note that gravity is applied by setting the external force load $\bar{\mathbf{f}}=-\rho Ag\{q_{1}q_3-q_0q_2 ,\ q_2 q_3+q_0 q_1,\ q_0^2+q_3^2-1/2\}$ in Eq.~\eqref{eqn:16}.
To achieve the desired geometry, we control the $z$-coordinate of the left end of the helix, $z(1.9)$, through an optimization procedure. The loss function is formulated as:
\begin{equation}
f=(z(1.9)-z_{tmp})^2.
\end{equation}

We gradually change $z_{tmp}$ from $R$ to $-R$, and shows height variance $\lambda$ in Fig.~\ref{fig:f10}(D) for $\alpha=10$. It shows that if there is no gravity, the minimal volume point is in $z_{tmp}=0$ and gravity causes the minimum point to shift to the right. As the gravity-elasticity number $\gamma$ gradually increases, the minimum value of the curve first decreases and then increases, which means that we can obtain the optimal value of $\lambda$ by adjusting the $w$ of the ribbon, thereby minimizing the volume of the antenna. For different width-to-thickness ratio $\alpha$, we plot the minimal $\lambda$ as a function of $\gamma$ in Fig.~\ref{fig:f10}(E), which can provide guidance for the engineering design of hemispherical helix antennas. We selected a copper antenna as an specific example, which means the density $\rho=8.96\ kg/m^3$, Young's modulus $E=120\ GPa$. Suppose the radius $R$ is 0.05 $m$ and the length of the ribbon is 0.922 $m$, the width-thickness ratio $\alpha$ is 10 and we can read the optimal $\gamma=0.078$ as shown in Fig.~\ref{fig:f10}(E) As a result, the fabrication width of the ribbon can be finally estimated $w=\sqrt{\rho g L^3/E\gamma}\approx2.7\ mm$.

\section{Discussion}
\label{sec:dis}
Building upon the theoretical framework presented in Section~\ref{sec: framework}, this study has established the inverse elastica theory as a direct solution method for determining UC from DC. The proposed methodology has demonstrated remarkable versatility through several examples: the inverse design of a trefoil knot and the discretization of helical curved surfaces with variable Gaussian curvatures.
The key contributions of this work can be systematically summarized as follows:
Firstly, we have formulated the inverse elastica theory as a novel approach for the inverse design of elastic rods. Comparative analysis reveals that the governing equations of inverse elastica exhibit significantly reduced nonlinearity compared to the forward Kirchhoff rod model.
Then our investigation elucidates the fundamental origin of multiple solutions in the inverse problem. These non-unique solutions arise from the inherent flexibility in specifying force and moment boundary conditions, which can be arbitrarily supplied through displacement constraints.
Moreover, by reformulating the inverse elastica equations as a BVP, we introduce the fundamental concept of inverse loading. This novel approach enables direct determination of UC from prescribed DC through controlled boundary displacements, thereby establishing a rigorous mathematical duality with the classical elastica theory. To address cases where desired UC characteristics cannot be conveniently expressed as BVP in Eq.~\eqref{eqn:16}, we further develop a theory-assisted optimization strategy within our theoretical framework.

Although the theoretical framework of inverse elastica has been established, several open questions remain for future investigation.
First, we are developing an inverse discrete elastic rod (InvDER) simulator capable of determining the UC from a given DC under prescribed inverse loading based on inverse elastica~\citep{lijh_InvDER2025}. This computational tool will serve as the inverse counterpart to the widely used DER simulator, which has found extensive applications in mechanics, computer graphics, biophysics, and related fields. For form-finding tasks involving more complex structures such as gridshells, which are often intractable by analytical means~\citep{baek2018form,qin2020genetic}, the InvDER framework holds great promise for enabling direct numerical solutions.
Second, similar to forward problems, the inverse loading framework exhibits bifurcation phenomena when analyzed using numerical continuation methods~\citep{yu2019bifurcations,yu2021numerical}. A comprehensive understanding of the physical significance of these bifurcations in the context of inverse problems has yet to be developed.
Third, inspired by Kirchhoff’s classical analogy between elastic rod theory and rigid body dynamics~\citep{kehrbaum1997elastic}, we plan to explore potential mathematical analogies for the inverse elastica formulation. Such analogies may offer deeper theoretical insights and unify inverse deformation mechanics with other areas of applied physics and geometry.
From an engineering perspective, we propose to relax the strict target shape constraints of conformable hemispherical helical ribbons in order to improve spatial efficiency. For instance, by optimizing the pitch distribution of the helix as DC, it may be possible to achieve reduced height variance and more compact deployable structures. Lastly, experimental fabrication of morphing hemispherical helical antennas designed via the inverse elastica theory is a crucial direction for future work.

\section{Conclusion}
\label{sec:conclusion}
We have presented inverse elastica theory as a general framework for determining the UC from a prescribed DC of slender structures in shape-morphing. By reformulating the inverse problem as a boundary value problem, the theory establishes a dual relationship with classical Kirchhoff theory and exhibits reduced nonlinearity compared to the forward elastica theory.
To clarify the core features of the framework, which include reduced nonlinearity, multiplicity of solutions, and the concept of inverse loading, we examined two representative examples: the inverse design of an arc and the inverse design of a helix. These cases were studied using both analytical solutions and numerical continuation methods. The theory was further validated through the inverse design of complex spatial configurations such as trefoil knots and helix discretized curved surfaces with varying Gaussian curvatures, validated with both DER simulations and experiments.
Looking ahead, future work includes the development of an inverse discrete elastic rod (InvDER) solver, the analysis of bifurcation behavior, and the fabrication of deployable and conformable hemisphere helix antennas. These directions will be essential for extending the applicability of the framework to fields such as flexible electronics, aerospace systems, and reconfigurable structures.
Inverse elastica theory provides a robust and theoretical foundation for the inverse design of morphing slender structures, with broad potential for applications in radio-frequency systems, deployable architecture, soft robotics and other related fields.

\section*{CRediT authorship contribution statement}
\textbf{Jiahao Li}: Conceptualization, Review \& Writing – original draft, Methodology, Software, Investigation, Data curation, Validation. \textbf{Weicheng Huang}: Review \& Writing – original draft, Methodology, Supervision, Formal analysis, Validation, Software.  \textbf{Yinbo Zhu}: Writing – Review \& Editing, Resources, Funding acquisition. \textbf{Luxia Yu}: Writing – Review \& Editing, Formal analysis. \textbf{Xiaohao Sun}: Writing – Review \& Editing, Formal analysis, Validation. \textbf{Mingchao Liu}: Conceptualization, Review \& Writing – original draft, Methodology, Supervision, Formal analysis,  Validation. \textbf{Hengan Wu}: Conceptualization, Writing – Review \& Editing, Validation, Supervision, Resources, Project administration, Funding acquisition.

\section*{Declaration of competing interest}
The authors declare that they have no known competing financial interests or personal relationships that could have appeared to influence the work reported in this paper.

\section*{Acknowledgements}

We are grateful to Haiyi Liang, Linghui He and Liu Wang at the University of Science and Technology of China for their insightful discussions. The research leading to these results has received funding from the National Natural Science Foundation of China (12388101, 12232016), the USTC Tang Scholar and the Fundamental Research Funds for the Central Universities (WK2090000087), and from the University of Birmingham via the start-up funding (M.L.). The numerical calculations were performed on the supercomputing system in Hefei Advanced Computing Center and the Supercomputing Center of University of Science and Technology of China.

\appendix
\renewcommand{\theequation}{\Alph{section}.\arabic{equation}} 
\section{The equations of helix discretized curved surfaces and experiment settings.}
\label{app:A}
\setcounter{equation}{0}
\noindent \textbf{The equations of helix discretized curved surfaces}

The equation of helix discretized cone is:

\begin{equation}
    \begin{cases}
	x =R\left( 1-t/2 \right) \cos \left( 2\pi Nt \right)\\
	y=R\left( 1-t/2 \right) \sin \left( 2\pi Nt \right)\\
	z=Rt\\
\end{cases}\ \ \ \ \ \ \ \ \ t\in \left[ 0,1.5 \right],
\end{equation}
where $R=0.05$,$N=2$.

The equation of helix discretized sphere is:

\begin{equation}
    \begin{cases}
	x=R\cos \left( 2\pi Nt \right) \sin \left( \pi /2t \right)\\
	y=R\sin \left( 2\pi Nt \right) \sin \left( \pi /2t \right)\\
	z=R\cos \left( \pi /2t \right)\\
\end{cases}\ \ \ \ \ \ \ \ \ t\in \left[ 0.1,1.9 \right],
\end{equation}
where $R=0.05$,$N=5$.

The equation of helix discretized hyperboloid is:

\begin{equation}
\begin{cases}
	x=R\left( \left( t-1 \right) ^2+0.5 \right) \cos \left( \pi Nt \right)\\
	y=R\left( \left( t-1 \right) ^2+0.5 \right) \sin \left( \pi Nt \right)\\
	z=tR/2\\
\end{cases}\ \ \ \ \ \ \ \ \ t\in \left[ 0,1.9 \right],
\end{equation}
where $R=0.05$,$N=5$.
\\

\noindent \textbf{Experiment settings}

The material properties employed in both the simulations and theoretical analysis are specified as follows: the Young's modulus is $2 \times 10^7\ \text{Pa}$, and the radius of the circular cross-section is $0.0015\ \text{m}$ (i.e., $1.5\ \text{mm}$).The STL models were built with OpenSCAD. The optimized helical structures, discretized into curved surface elements, were fabricated using a Bamboo X1C 3D printer with SLA resin. These slender SLA-printed structures are capable of sustaining small strains while exhibiting significant rotational deformation behavior. The effect of gravity is not considered in these three cases.

\section{The discrete elastic rod method.}
\label{app:B}
\setcounter{equation}{0}
\noindent \textbf{The loading steps of DER verifications}

To validate the proposed inverse elastica theory, simulations based on the Discrete Elastic Rod (DER) are conducted~\citep{bergou2008discrete,huang2025tutorial,jawed2018primer}. The loading procedure consists of the following steps: first, the positions of two nodes at the left end of the UC are fixed. Next, the positions of two nodes at the right end of the UC are gradually displaced to match those of the corresponding nodes at the right end of DC. Finally, the material director at the terminal edge of the UC is rotated to align with that of the DC.
\\

\noindent \textbf{Discrete elastic rod algorithm}
\begin{figure}[ht]
    \centering
    \includegraphics[width=0.6\columnwidth]{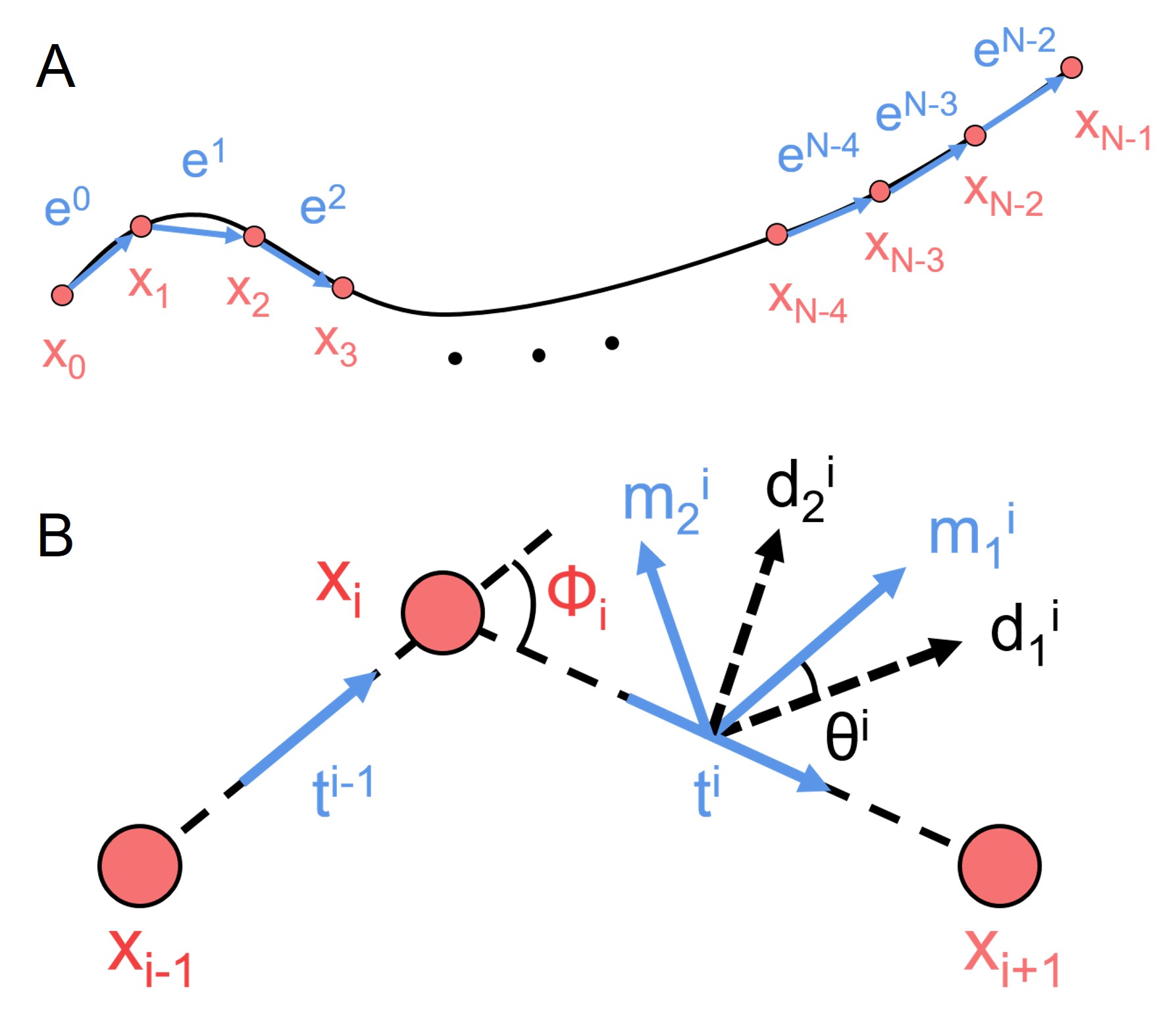}
    \caption{\textbf{The schematic diagram of discrete elastic rod algorithm.} (A) The slender structure as a discrete cosserate curve. (B) The bending element and the associated material frame and reference frame.}
    \label{fig:f11}
\end{figure}
The discrete elastica rod (DER) method is used to validate our inverse elastica theory by forward loading from the UC to the DC~\citep{bergou2008discrete,huang2025tutorial,jawed2018primer}. As shown in Fig.~\ref{fig:f11}(A), a slender structure is discretized into $N$ nodes $ \boldsymbol{x}_0, \boldsymbol{x}_1, ..., \boldsymbol{x}_{N-1} $, forming $N - 1$ edge vectors $ \boldsymbol{e}^i = \boldsymbol{x}_{i+1} - \boldsymbol{x}_i $ for $ i = 0, ..., N - 2 $. Subscripts denote node-based quantities and superscripts denote edge-based quantities. Each edge $ \boldsymbol{e}^i $ has an orthonormal reference frame $ \{ \mathbf{d}_1^i, \mathbf{d}_2^i, \boldsymbol{t}^i \} $ and a material frame $ \{ \boldsymbol{m}_1^i, \boldsymbol{m}_2^i, \boldsymbol{t}^i \} $, sharing the tangent $ \boldsymbol{t}^i = \boldsymbol{e}^i / \| \boldsymbol{e}^i \| $. The reference frame is updated via parallel transport, and the material frame is obtained by applying a twist angle $ \theta^i $ (Fig.~\ref{fig:f11}(B)). The generalized coordinate vector is
$ \boldsymbol{q} = [\boldsymbol{x}_0, \theta^0, \boldsymbol{x}_1, \theta^1, ..., \boldsymbol{x}_{N-2}, \theta^{N-2}, \boldsymbol{x}_{N-1}]^T $,
with $ 4N - 1 $ degrees of freedom (DOF), where $ (\cdot)^T $ denotes vector transpose.

\textbf{Strain definitions:}

\textit{Stretching strain:}
\begin{equation}
    \varepsilon^i = \frac{\| \boldsymbol{e}^i \|}{\| \bar{\boldsymbol{e}}^i \|} - 1
\end{equation}
where $\| \cdot \|$ is the norm, and $ \bar{\boldsymbol{e}}^i $ is the reference edge vector.

\textit{Bending strain:} The Darboux binormal vector is
\begin{equation}
    (\kappa \boldsymbol{b})_i = \frac{2 \boldsymbol{e}^{i-1} \times \boldsymbol{e}^i}{\| \boldsymbol{e}^{i-1} \| \| \boldsymbol{e}^i \| + \boldsymbol{e}^{i-1} \cdot \boldsymbol{e}^i}
\end{equation}
with magnitude $ \| (\kappa \boldsymbol{b})_i \| = 2 \tan(\phi_i / 2) $.

Material curvatures can be calculated as bending strain:
\begin{equation}
    \kappa_i^{(1)} = \frac{1}{2} (\boldsymbol{m}_2^{i-1} + \boldsymbol{m}_2^i) \cdot (\kappa \boldsymbol{b})_i
\end{equation}
\begin{equation}
    \kappa_i^{(2)} = -\frac{1}{2} (\boldsymbol{m}_1^{i-1} + \boldsymbol{m}_1^i) \cdot (\kappa \boldsymbol{b})_i
\end{equation}

\textit{Twisting strain:}
\begin{equation}
    \tau_i = \theta^i - \theta^{i-1} + m_i
    \label{eqn:twist}
\end{equation}
where $ m_i $ is the reference twist. Eq.~\eqref{eqn:twist} expresses the total twist as the sum of reference twist and relative twist.

\textbf{Elastic energies:}

\textit{Stretching energy:}
\begin{equation}
    E_s = \frac{1}{2} \sum_{i=0}^{N-2} EA (\varepsilon^i)^2 \| \bar{\boldsymbol{e}}^i \|
\end{equation}

\textit{Bending energy:}
\begin{equation}
    E_b = \frac{1}{2} \sum_{i=1}^{N-2} \frac{1}{\Delta l_i} \left[ EI_1 (\kappa_i^{(1)} - \bar{\kappa}_i^{(1)})^2 + EI_2 (\kappa_i^{(2)} - \bar{\kappa}_i^{(2)})^2 \right]
\end{equation}

\textit{Twisting energy:}
\begin{equation}
    E_t = \frac{1}{2} \sum_{i=1}^{N-2} \frac{GJ}{\Delta l_i} (\tau_i)^2
\end{equation}
where $ \Delta l_i $ is the Voronoi length at node $ \boldsymbol{x}_i $.

\textbf{Time integration (Implicit Euler):}

Let $ \boldsymbol{v} = \dot{\boldsymbol{q}} $. The update from $ t_k $ to $ t_{k+1} = t_k + h $ follows:
\begin{equation}
    \mathbb{M} (\boldsymbol{q}_{k+1} - \boldsymbol{q}_k - h \boldsymbol{v}_k) - h^2 (\boldsymbol{F}_{k+1}^{\mathrm{int}} + \boldsymbol{F}_{k+1}^{\mathrm{ext}}) = \boldsymbol{0}
    \label{eqn:newton}
\end{equation}
where $ \boldsymbol{F}_{k+1}^{\mathrm{int}} = -\frac{\partial}{\partial \boldsymbol{q}_{k+1}} (E_s + E_b + E_t) $, and $ \boldsymbol{F}_{k+1}^{\mathrm{ext}} $ is the external force (e.g., gravity). $ \mathbb{M} $ is the lumped mass matrix.

The Jacobian matrix for Eq.~\eqref{eqn:newton} is
\begin{equation}
    \mathbb{J}_{ij} = \mathbb{M}_{ij} - h^2 \left( - \frac{\partial^2 (E_s + E_b + E_t)}{\partial q_i \partial q_j} \right)
\end{equation}
where $ q_i $ is the $ i $-th component of $ \boldsymbol{q} $. The Jacobian $ \mathbb{J} $ is banded, and the time complexity is $ \mathcal{O}(N) $.

\section{Vedio of DER verification}
\label{app:C}
The forward verification for the elastic deformation of trefoil knot, helix discretized torus, cone, sphere and hyperboloid surfaces from UC to DC.

\bibliographystyle{apalike} 

\bibliography{Bibliography}

\begin{thebibliography}{}

\bibitem[Abbena et~al., 2017]{abbena2017modern}
Abbena, E., Salamon, S., and Gray, A. (2017).
\newblock {\em Modern differential geometry of curves and surfaces with Mathematica}.
\newblock Chapman and Hall/CRC.

\bibitem[Aharoni et~al., 2018]{aharoni2018universal}
Aharoni, H., Xia, Y., Zhang, X., Kamien, R.~D., and Yang, S. (2018).
\newblock Universal inverse design of surfaces with thin nematic elastomer sheets.
\newblock {\em Proceedings of the National Academy of Sciences}, 115(28):7206--7211.

\bibitem[Armon et~al., 2011]{armon2011geometry}
Armon, S., Efrati, E., Kupferman, R., and Sharon, E. (2011).
\newblock Geometry and mechanics in the opening of chiral seed pods.
\newblock {\em Science}, 333(6050):1726--1730.

\bibitem[Audoly and Pomeau, 2000]{audoly2000elasticity}
Audoly, B. and Pomeau, Y. (2000).
\newblock Elasticity and geometry.
\newblock In {\em Peyresq lectures on nonlinear phenomena}, pages 1--35. World Scientific.

\bibitem[Baek et~al., 2018]{baek2018form}
Baek, C., Sageman-Furnas, A.~O., Jawed, M.~K., and Reis, P.~M. (2018).
\newblock Form finding in elastic gridshells.
\newblock {\em Proceedings of the National Academy of Sciences}, 115(1):75--80.

\bibitem[Benvenuto et~al., 2015]{benvenuto2015dynamics}
Benvenuto, R., Salvi, S., and Lavagna, M. (2015).
\newblock Dynamics analysis and gnc design of flexible systems for space debris active removal.
\newblock {\em Acta Astronautica}, 110:247--265.

\bibitem[Bergou et~al., 2008]{bergou2008discrete}
Bergou, M., Wardetzky, M., Robinson, S., Audoly, B., and Grinspun, E. (2008).
\newblock Discrete elastic rods.
\newblock In {\em ACM SIGGRAPH 2008 Papers}, pages 1--12.

\bibitem[Bertails-Descoubes et~al., 2018]{bertails2018inverse}
Bertails-Descoubes, F., Derouet-Jourdan, A., Romero, V., and Lazarus, A. (2018).
\newblock Inverse design of an isotropic suspended kirchhoff rod: theoretical and numerical results on the uniqueness of the natural shape.
\newblock {\em Proceedings of the Royal Society A: Mathematical, Physical and Engineering Sciences}, 474(2212):20170837.

\bibitem[Boley et~al., 2019]{boley2019shape}
Boley, J.~W., Van~Rees, W.~M., Lissandrello, C., Horenstein, M.~N., Truby, R.~L., Kotikian, A., Lewis, J.~A., and Mahadevan, L. (2019).
\newblock Shape-shifting structured lattices via multimaterial 4d printing.
\newblock {\em Proceedings of the National Academy of Sciences}, 116(42):20856--20862.

\bibitem[Chen et~al., 2011]{chen2011tunable}
Chen, Z., Majidi, C., Srolovitz, D.~J., and Haataja, M. (2011).
\newblock Tunable helical ribbons.
\newblock {\em Applied Physics Letters}, 98(1).

\bibitem[Cheng et~al., 2023]{cheng2023programming}
Cheng, X., Fan, Z., Yao, S., Jin, T., Lv, Z., Lan, Y., Bo, R., Chen, Y., Zhang, F., Shen, Z., et~al. (2023).
\newblock Programming 3d curved mesosurfaces using microlattice designs.
\newblock {\em Science}, 379(6638):1225--1232.

\bibitem[Chu, 1948]{chu1948physical}
Chu, L.~J. (1948).
\newblock Physical limitations of omni-directional antennas.
\newblock {\em Journal of Applied Physics}, 19(12):1163--1175.

\bibitem[Derouet-Jourdan et~al., 2013]{derouet2013inverse}
Derouet-Jourdan, A., Bertails-Descoubes, F., Daviet, G., and Thollot, J. (2013).
\newblock Inverse dynamic hair modeling with frictional contact.
\newblock {\em ACM Transactions on Graphics (TOG)}, 32(6):1--10.

\bibitem[Derouet-Jourdan et~al., 2010]{derouet2010stable}
Derouet-Jourdan, A., Bertails-Descoubes, F., and Thollot, J. (2010).
\newblock Stable inverse dynamic curves.
\newblock {\em ACM Transactions on Graphics (TOG)}, 29(6):1--10.

\bibitem[Do~Carmo, 2016]{do2016differential}
Do~Carmo, M.~P. (2016).
\newblock {\em Differential geometry of curves and surfaces: revised and updated second edition}.
\newblock Courier Dover Publications.

\bibitem[Doedel et~al., 2007]{doedel2007auto}
Doedel, E.~J., Champneys, A.~R., Dercole, F., Fairgrieve, T.~F., Kuznetsov, Y.~A., Oldeman, B., Paffenroth, R., Sandstede, B., Wang, X., and Zhang, C. (2007).
\newblock Auto-07p: Continuation and bifurcation software for ordinary differential equations.

\bibitem[Efrati et~al., 2009]{efrati2009elastic}
Efrati, E., Sharon, E., and Kupferman, R. (2009).
\newblock Elastic theory of unconstrained non-euclidean plates.
\newblock {\em Journal of the Mechanics and Physics of Solids}, 57(4):762--775.

\bibitem[Fan et~al., 2020]{fan2020inverse}
Fan, Z., Yang, Y., Zhang, F., Xu, Z., Zhao, H., Wang, T., Song, H., Huang, Y., Rogers, J.~A., and Zhang, Y. (2020).
\newblock Inverse design strategies for 3d surfaces formed by mechanically guided assembly.
\newblock {\em Advanced Materials}, 32(14):1908424.

\bibitem[Huang et~al., 2025]{huang2025tutorial}
Huang, W., Hao, Z., Li, J., Tong, D., Guo, K., Zhang, Y., Gao, H., Hsia, K.~J., and Liu, M. (2025).
\newblock A tutorial on simulating nonlinear behaviors of flexible structures with the discrete differential geometry (ddg) method.
\newblock {\em Applied Mechanics Reviews}, pages 1--88.

\bibitem[Huang et~al., 2023]{huang2023contact}
Huang, W., Zou, H., Liu, H., Yang, W., Gao, J., and Liu, Z. (2023).
\newblock Contact dynamic analysis of tether-net system for space debris capture using incremental potential formulation.
\newblock {\em Advances in Space Research}, 72(6):2039--2050.

\bibitem[Jawed et~al., 2018]{jawed2018primer}
Jawed, M.~K., Novelia, A., and O'Reilly, O.~M. (2018).
\newblock {\em A primer on the kinematics of discrete elastic rods}.
\newblock Springer.

\bibitem[Kansara et~al., 2023]{kansara2023inverse}
Kansara, H., Liu, M., He, Y., and Tan, W. (2023).
\newblock Inverse design and additive manufacturing of shape-morphing structures based on functionally graded composites.
\newblock {\em Journal of the Mechanics and Physics of Solids}, 180:105382.

\bibitem[Kehrbaum and Maddocks, 1997]{kehrbaum1997elastic}
Kehrbaum, S. and Maddocks, J. (1997).
\newblock Elastic rods, rigid bodies, quaternions and the last quadrature.
\newblock {\em Philosophical Transactions of the Royal Society of London. Series A: Mathematical, Physical and Engineering Sciences}, 355(1732):2117--2136.

\bibitem[Kim et~al., 2019]{kim2019ferromagnetic}
Kim, Y., Parada, G.~A., Liu, S., and Zhao, X. (2019).
\newblock Ferromagnetic soft continuum robots.
\newblock {\em Science Robotics}, 4(33):eaax7329.

\bibitem[Kong et~al., 2016]{kong2016electrically}
Kong, M., Shin, G., Lee, S.-H., and Yoon, I.-J. (2016).
\newblock Electrically small folded spherical helix antennas using copper strips and 3d printing technology.
\newblock {\em Electronics Letters}, 52(12):994--996.

\bibitem[Lee et~al., 2020]{lee2020computational}
Lee, Y.-K., Xi, Z., Lee, Y.-J., Kim, Y.-H., Hao, Y., Choi, H., Lee, M.-G., Joo, Y.-C., Kim, C., Lien, J.-M., et~al. (2020).
\newblock Computational wrapping: A universal method to wrap 3d-curved surfaces with nonstretchable materials for conformal devices.
\newblock {\em Science Advances}, 6(15):eaax6212.

\bibitem[Levien, 2008]{levien2008elastica}
Levien, R. (2008).
\newblock The elastica: a mathematical history.
\newblock Technical report, Technical Report No. UCB/EECS-2008-103.

\bibitem[Li, 2025]{lijh_InvDER2025}
Li, J. (2025).
\newblock The inverse discrete elastic rod.
\newblock {\em To be submitted}.

\bibitem[Li et~al., 2025a]{li2025biomimetic}
Li, J., Sun, X., He, Z., Hou, Y., Wu, H., and Zhu, Y. (2025a).
\newblock Biomimetic turing machine: A multiscale theoretical framework for the inverse design of target space curves.
\newblock {\em Journal of the Mechanics and Physics of Solids}, 196:105999.

\bibitem[Li et~al., 2025b]{li2025harnessing}
Li, J., Tong, D., Hao, Z., Zhu, Y., Wu, H., Liu, M., and Huang, W. (2025b).
\newblock Harnessing discrete differential geometry: A virtual playground for the bilayer soft robotics.
\newblock {\em Advanced Intelligent Systems}, page 2500141.

\bibitem[Ling et~al., 2024]{ling2024tension}
Ling, S., Tian, X., Zeng, Q., Qin, Z., Kurt, S.~A., Tan, Y.~J., Fuh, J.~Y., Liu, Z., Dickey, M.~D., Ho, J.~S., et~al. (2024).
\newblock Tension-driven three-dimensional printing of free-standing field’s metal structures.
\newblock {\em Nature Electronics}, 7(8):671--683.

\bibitem[Liu et~al., 2020]{liu2020tapered}
Liu, M., Domino, L., and Vella, D. (2020).
\newblock Tapered elastic{\ae} as a route for axisymmetric morphing structures.
\newblock {\em Soft Matter}, 16(33):7739--7750.

\bibitem[Love, 1944]{love1944treatise}
Love, A. E.~H. (1944).
\newblock {\em A treatise on the mathematical theory of elasticity}.
\newblock Courier Corporation.

\bibitem[Matsutani, 2010]{matsutani2010euler}
Matsutani, S. (2010).
\newblock Euler's elastica and beyond.
\newblock {\em Journal of Geometry and Symmetry in Physics}, 17:45--86.

\bibitem[Matsutani, 2024]{matsutani2024euler}
Matsutani, S. (2024).
\newblock Euler's original derivation of elastica equation.
\newblock {\em arXiv preprint arXiv:2411.09227}.

\bibitem[Moulton et~al., 2018]{moulton2018stable}
Moulton, D.~E., Grandgeorge, P., and Neukirch, S. (2018).
\newblock Stable elastic knots with no self-contact.
\newblock {\em Journal of the Mechanics and Physics of Solids}, 116:33--53.

\bibitem[O'Reilly, 2017]{o2017modeling}
O'Reilly, O.~M. (2017).
\newblock {\em Modeling nonlinear problems in the mechanics of strings and rods}.
\newblock Springer.

\bibitem[Qin et~al., 2020]{qin2020genetic}
Qin, L., Huang, W., Du, Y., Zheng, L., and Jawed, M.~K. (2020).
\newblock Genetic algorithm-based inverse design of elastic gridshells.
\newblock {\em Structural and Multidisciplinary Optimization}, 62:2691--2707.

\bibitem[Qin et~al., 2022]{qin2022bottom}
Qin, L., Zhu, J., and Huang, W. (2022).
\newblock A bottom-up optimization method for inverse design of two-dimensional clamped-free elastic rods.
\newblock {\em International Journal for Numerical Methods in Engineering}, 123(11):2556--2572.

\bibitem[Rushen, 2019]{rushen2019viper}
Rushen (2019).
\newblock Trimeresurus sabahi fucatus, banded pit viper - takua pa district, phang-nga province.
\newblock \url{https://commons.wikimedia.org/wiki/File:Trimeresurus_sabahi_fucatus,_Banded_pit_viper_-_Takua_Pa_District,_Phang-nga_Province_(46710893582).jpg}.
\newblock Image licensed under CC BY 2.0 via Wikimedia Commons.

\bibitem[Sawa et~al., 2011]{sawa2011shape}
Sawa, Y., Ye, F., Urayama, K., Takigawa, T., Gimenez-Pinto, V., Selinger, R.~L., and Selinger, J.~V. (2011).
\newblock Shape selection of twist-nematic-elastomer ribbons.
\newblock {\em Proceedings of the National Academy of Sciences}, 108(16):6364--6368.

\bibitem[Shi et~al., 2025]{shi2025double}
Shi, Q., Huang, W., Yu, T., and Li, M. (2025).
\newblock Double-eigenvalue bifurcation and multistability in serpentine strips with tunable buckling behaviors.
\newblock {\em Journal of the Mechanics and Physics of Solids}, 195:105922.

\bibitem[Shin et~al., 2018]{shin2018hygrobot}
Shin, B., Ha, J., Lee, M., Park, K., Park, G.~H., Choi, T.~H., Cho, K.-J., and Kim, H.-Y. (2018).
\newblock Hygrobot: A self-locomotive ratcheted actuator powered by environmental humidity.
\newblock {\em Science Robotics}, 3(14):eaar2629.

\bibitem[Si{\'e}fert et~al., 2019]{siefert2019bio}
Si{\'e}fert, E., Reyssat, E., Bico, J., and Roman, B. (2019).
\newblock Bio-inspired pneumatic shape-morphing elastomers.
\newblock {\em Nature Materials}, 18(1):24--28.

\bibitem[Singer and Nelder, 2009]{singer2009nelder}
Singer, S. and Nelder, J. (2009).
\newblock Nelder-mead algorithm.
\newblock {\em Scholarpedia}, 4(7):2928.

\bibitem[Sun et~al., 2022]{sun2022phase}
Sun, X., Wu, S., Dai, J., Leanza, S., Yue, L., Yu, L., Jin, Y., Qi, H.~J., and Zhao, R.~R. (2022).
\newblock Phase diagram and mechanics of snap-folding of ring origami by twisting.
\newblock {\em International Journal of Solids and Structures}, 248:111685.

\bibitem[Sydney~Gladman et~al., 2016]{sydney2016biomimetic}
Sydney~Gladman, A., Matsumoto, E.~A., Nuzzo, R.~G., Mahadevan, L., and Lewis, J.~A. (2016).
\newblock Biomimetic 4d printing.
\newblock {\em Nature Materials}, 15(4):413--418.

\bibitem[Timoshenko, 1925]{timoshenko1925analysis}
Timoshenko, S. (1925).
\newblock Analysis of bi-metal thermostats.
\newblock {\em Journal of the Optical Society of America}, 11(3):233--255.

\bibitem[Tong et~al., 2025]{tong2025inverse}
Tong, D., Hao, Z., Li, J., and Huang, W. (2025).
\newblock Inverse design of planar clamped-free elastic rods from noisy data.
\newblock {\em International Journal for Numerical Methods in Engineering}, 126(5):e70018.

\bibitem[Van~Rees et~al., 2017]{van2017growth}
Van~Rees, W.~M., Vouga, E., and Mahadevan, L. (2017).
\newblock Growth patterns for shape-shifting elastic bilayers.
\newblock {\em Proceedings of the National Academy of Sciences}, 114(44):11597--11602.

\bibitem[Wang et~al., 2023]{wang2023curvature}
Wang, T., Dai, Z., Potier-Ferry, M., and Xu, F. (2023).
\newblock Curvature-regulated multiphase patterns in tori.
\newblock {\em Physical Review Letters}, 130(4):048201.

\bibitem[Wang et~al., 2025]{wang2025nonlinear}
Wang, T., Potier-Ferry, M., and Xu, F. (2025).
\newblock A nonlinear toroidal shell model for surface morphologies and morphogenesis.
\newblock {\em Journal of the Mechanics and Physics of Solids}, 200:106135.

\bibitem[Xu et~al., 2015]{xu2015assembly}
Xu, S., Yan, Z., Jang, K.-I., Huang, W., Fu, H., Kim, J., Wei, Z., Flavin, M., McCracken, J., Wang, R., et~al. (2015).
\newblock Assembly of micro/nanomaterials into complex, three-dimensional architectures by compressive buckling.
\newblock {\em Science}, 347(6218):154--159.

\bibitem[Yang et~al., 2023]{yang2023morphing}
Yang, X., Zhou, Y., Zhao, H., Huang, W., Wang, Y., Hsia, K.~J., and Liu, M. (2023).
\newblock Morphing matter: From mechanical principles to robotic applications.
\newblock {\em Soft Science}, 3(4):38.

\bibitem[Yu et~al., 2021]{yu2021numerical}
Yu, T., Dreier, L., Marmo, F., Gabriele, S., Parascho, S., and Adriaenssens, S. (2021).
\newblock Numerical modeling of static equilibria and bifurcations in bigons and bigon rings.
\newblock {\em Journal of the Mechanics and Physics of Solids}, 152:104459.

\bibitem[Yu and Hanna, 2019]{yu2019bifurcations}
Yu, T. and Hanna, J. (2019).
\newblock Bifurcations of buckled, clamped anisotropic rods and thin bands under lateral end translations.
\newblock {\em Journal of the Mechanics and Physics of Solids}, 122:657--685.

\bibitem[Yu et~al., 2023]{yu2023continuous}
Yu, T., Marmo, F., Cesarano, P., and Adriaenssens, S. (2023).
\newblock Continuous modeling of creased annuli with tunable bistable and looping behaviors.
\newblock {\em Proceedings of the National Academy of Sciences}, 120(4):e2209048120.

\bibitem[Zhang et~al., 2022]{zhang2022shape}
Zhang, Y., Yang, J., Liu, M., and Vella, D. (2022).
\newblock Shape-morphing structures based on perforated kirigami.
\newblock {\em Extreme Mechanics Letters}, 56:101857.

\end{thebibliography}

\end{document}